\documentclass[a4paper,11pt]{article}
\pdfoutput=1 % if your are submitting a pdflatex (i.e. if you have
\usepackage[separate-uncertainty,retain-explicit-plus,per-mode=symbol,binary-units]{siunitx}
\usepackage[export]{adjustbox}
\usepackage{wrapfig}
\usepackage{ragged2e}
\usepackage{array,mathtools,dcolumn}
\usepackage{amsmath,amssymb}
\usepackage{wasysym}
\usepackage{stmaryrd}
\usepackage[below]{placeins}
\usepackage[table]{xcolor}
\usepackage{soul}
\usepackage{tikz}
\usepackage{afterpage}
\usepackage{lineno}
\usepackage{listings}
\usepackage{array}
\usepackage[version=4]{mhchem}
\usepackage{multirow}
\usepackage{eurosym}
\usepackage{pagecolor}
\usepackage{fancyhdr}
\usepackage{etoolbox}
\usepackage{calc}
\usepackage{dcolumn}

\usepackage{bm}
\usepackage{xargs}
\usepackage{xfrac}
\usepackage[colorinlistoftodos,prependcaption,textsize=small]{todonotes}
\newcommandx{\unsure}[2][1=]{\todo[linecolor=red,backgroundcolor=red!25,bordercolor=red,#1]{#2}}
\newcommandx{\change}[2][1=]{\todo[linecolor=blue,backgroundcolor=blue!25,bordercolor=blue,#1]{#2}}
\newcommandx{\info}[2][1=]{\todo[linecolor=OliveGreen,backgroundcolor=OliveGreen!25,bordercolor=OliveGreen,#1]{#2}}
\newcommandx{\improvement}[2][1=]{\todo[linecolor=Plum,backgroundcolor=Plum!25,bordercolor=Plum,#1]{#2}}
\newcommandx{\thiswillnotshow}[2][1=]{\todo[disable,#1]{#2}}

\newcommand{\reffig}[1]{Fig.~\ref{fig:#1}}

\usepackage{paralist}
\usepackage{tcolorbox}
\usepackage{appendix}
\usepackage{enumitem}
\setdescription{itemsep=0pt,parsep=0pt,labelsep=.2cm,leftmargin=2.2cm,labelwidth=2cm,listparindent=.7cm}
\setlist{nolistsep,leftmargin=1cm}
\newlist{enumcompactitem}{itemize}{3}
\setlist[enumcompactitem]{topsep=0pt,partopsep=0pt,itemsep=0pt,parsep=0pt}
\setlist[enumcompactitem,1]{label=\textbullet}
\setlist[enumcompactitem,2]{label=---}
\setlist[enumcompactitem,3]{label=*}
\newlist{enumcompactdesc}{description}{3}
\setlist[enumcompactdesc]{topsep=0pt,partopsep=0pt,itemsep=0pt,parsep=0pt}
\newlist{enumcompactenum}{enumerate}{3}
\setlist[enumcompactenum]{topsep=0pt,partopsep=0pt,itemsep=0pt,parsep=0pt}
\setlist[enumcompactenum,1]{label=\arabic*}
\setlist[enumcompactenum,2]{label=\alph*}
\setlist[enumcompactenum,3]{label=\roman*}
\DeclareSIUnit\c{\mbox{$c$}}
\DeclareSIUnit\magn{\mbox{$\times$}}
\DeclareSIUnit\min{min}
\DeclareSIUnit\week{week}
\DeclareSIUnit\month{mo}
\DeclareSIUnit\months{mos}
\DeclareSIUnit\year{yr}
\DeclareSIUnit\years{years}
\DeclareSIUnit\yr{yr}
\DeclareSIUnit\standard{std}
\DeclareSIUnit\str{sr}
\DeclareSIUnit\ppm{ppm}
\DeclareSIUnit\ppb{ppb}
\DeclareSIUnit\ppt{ppt}
\DeclareSIUnit\pe{PE}
\DeclareSIUnit\spe{SPE}
\DeclareSIUnit\pdm{PDM}
\DeclareSIUnit\ev{events}
\DeclareSIUnit\ct{counts}
\DeclareSIUnit\neutron{\mbox{$n$}}
\DeclareSIUnit\smp{samples}
\DeclareSIUnit\Sample{S}
\DeclareSIUnit\ch{ch}
\DeclareSIUnit\hit{hit}
\DeclareSIUnit\hits{hits}
\DeclareSIUnit\bin{(\mbox{5-PE}~bin)}
\DeclareSIUnit\sgm{\mbox{$\sigma$}}
\DeclareSIUnit\rms{RMS}
\DeclareSIUnit\keVee{\mbox{keV$_{e{\rm e}}$}}
\DeclareSIUnit\keVr{\mbox{keV$_{\rm nr}$}}
\DeclareSIUnit\eVee{\mbox{eV$_{\rm ee}$}}
\DeclareSIUnit\eVr{\mbox{eV$_{\rm nr}$}}
\DeclareSIUnit\ph{photon}
\DeclareSIUnit\el{\mbox{$e^-$}}
\DeclareSIUnit\pm{\mbox{PMT}}
\DeclareSIUnit\pixel{\mbox{pixel}}
\DeclareSIUnit\inch{''}
\DeclareSIUnit\foot{'}
\DeclareSIUnit\bit{bit}
\DeclareSIUnit\sample{samples}
\DeclareSIUnit\barn{barn}
\DeclareSIUnit\bara{bar}
\DeclareSIUnit\barg{barg}
\DeclareSIUnit\mlardepth{\mbox(meter~of~\LAr~depth)}
\DeclareSIUnit\Curie{Ci}
\DeclareSIUnit\psi{psi}
\DeclareSIUnit\psf{psf}
\DeclareSIUnit\pcf{pcf}
\DeclareSIUnit\parsec{pc}
\DeclareSIUnit\mwe{\mbox{m.w.e.}}
\DeclareSIUnit\liveday{\mbox{live-days}}
\DeclareSIUnit\days{\mbox{days}}
\DeclareSIUnit\miles{\mbox{miles}}
\DeclareSIUnit\lumens{\mbox{lm}}
\DeclareSIUnit\degreeC{\mbox{$^{\circ}$C}}
\DeclareSIUnit\degreeF{\mbox{$^{\circ}$F}}
\DeclareSIUnit\electron{\mbox{$e^-$}}
\DeclareSIUnit\Euro{\mbox{\euro}}
\DeclareSIUnit\cph{cph}
\DeclareSIUnit\neq{neq}
\DeclareSIUnit\normal{\mbox{N}}
\newcommand{\CEnNS}{\mbox{CE$\nu$NS}}
\newcommand{\DSf}{\mbox{DarkSide-50}}

\newcommand{\DSk}{\mbox{DarkSide-20k}}

\newcommand{\Argo}{\mbox{Argo}}

\newcommand{\Aria}{\mbox{Aria}}

\newcommand{\LAr}{\ce{LAr}}

\newcommand{\DSkSupernovaBaseline}{\SI{10}{\kilo\parsec}}

\newcommand{\ArThirtyNine}{\ce{^39Ar}}

\input{dsnew.def}
\input{SN.def}
\renewcommand{\reffig}[1]{figure~\ref{fig:#1}}
\usepackage{float}
\usepackage{jcappub} % for details on the use of the package, please
\usepackage[T1]{fontenc} % if needed

\usepackage[table]{xcolor}

\usepackage{lineno} 
\usepackage[normalem]{ulem}

\newcommand{\Alberta}{Department of Physics, University of Alberta, Edmonton, AB T6G 2R3, Canada}
\newcommand{\APC}{APC, Universit\'e de Paris, CNRS, Astroparticule et Cosmologie, Paris F-75013, France}
\newcommand{\AQLNGS}{INFN Laboratori Nazionali del Gran Sasso, Assergi (AQ) 67100, Italy}
\newcommand{\AQGSSI}{Gran Sasso Science Institute, L'Aquila 67100, Italy}

\newcommand{\AstroCeNT}{AstroCeNT, Nicolaus Copernicus Astronomical Center of the Polish Academy of Sciences, 00-614 Warsaw, Poland}
\newcommand{\Augustana}{Physics Department, Augustana University, Sioux Falls, SD 57197, USA}
\newcommand{\Belgorod}{Radiation Physics Laboratory, Belgorod National Research University, Belgorod 308007, Russia}

\newcommand{\BINP}{Budker Institute of Nuclear Physics, Novosibirsk 630090, Russia}
\newcommand{\BNLaddress}{Brookhaven National Laboratory, Upton, NY 11973, USA}
\newcommand{\BOINFN}{INFN Bologna, Bologna 40126, Italy}
\newcommand{\BOUniPHY}{Physics Department, Universit\`a degli Studi di Bologna, Bologna 40126, Italy}

\newcommand{\CAUniEEE}{Department of Electrical and Electronic Engineering, Universit\`a degli Studi di Cagliari, Cagliari 09123, Italy}
\newcommand{\CAUniPHY}{Physics Department, Universit\`a degli Studi di Cagliari, Cagliari 09042, Italy}
\newcommand{\CAINFN}{INFN Cagliari, Cagliari 09042, Italy}
\newcommand{\Carleton}{Department of Physics, Carleton University, Ottawa, ON K1S 5B6, Canada}

\newcommand{\CentroFermi}{Museo della fisica e Centro studi e Ricerche Enrico Fermi, Roma 00184, Italy}
\newcommand{\CERNaddress}{CERN, European Organization for Nuclear Research 1211 Geneve 23, Switzerland, CERN}
\newcommand{\CIEMAT}{CIEMAT, Centro de Investigaciones Energ\'eticas, Medioambientales y Tecnol\'ogicas, Madrid 28040, Spain}

\newcommand{\CPPM}{Centre de Physique des Particules de Marseille, Aix Marseille Univ, CNRS/IN2P3, CPPM, Marseille, France}
\newcommand{\CTINFN}{INFN Catania, Catania 95121, Italy}
\newcommand{\CTUNI}{Universit\`a of Catania, Catania 95124, Italy}
\newcommand{\CTLNS}{INFN Laboratori Nazionali del Sud, Catania 95123, Italy}
\newcommand{\ENUniCEE}{Engineering and Architecture Faculty, Universit\`a di Enna Kore, Enna 94100, Italy}
\newcommand{\ETHZ}{Institute for Particle Physics, ETH Z\"urich, Z\"urich 8093, Switzerland}
\newcommand{\FNALaddress}{Fermi National Accelerator Laboratory, Batavia, IL 60510, USA}
\newcommand{\FortLewis}{Department of Physics and Engineering, Fort Lewis College, Durango, CO 81301, USA}
\newcommand{\GEUni}{Physics Department, Universit\`a degli Studi di Genova, Genova 16146, Italy}
\newcommand{\GEINFN}{INFN Genova, Genova 16146, Italy}
\newcommand{\Hawaii}{Department of Physics and Astronomy, University of Hawai'i, Honolulu, HI 96822, USA}
\newcommand{\Houston}{Department of Physics, University of Houston, Houston, TX 77204, USA}
\newcommand{\IHEPaddress}{Institute of High Energy Physics, Beijing 100049, China}

\newcommand{\JINR}{Joint Institute for Nuclear Research, Dubna 141980, Russia}
\newcommand{\Krakow}{M.~Smoluchowski Institute of Physics, Jagiellonian University, 30-348 Krakow, Poland}
\newcommand{\Kurchatov}{National Research Centre Kurchatov Institute, Moscow 123182, Russia}
\newcommand{\Laurentian}{Department of Physics and Astronomy, Laurentian University, Sudbury, ON P3E 2C6, Canada}
\newcommand{\Lancaster}{Physics Department, Lancaster University, Lancaster LA1 4YB, UK}
\newcommand{\LNFINFN}{INFN Laboratori Nazionali di Frascati, Frascati 00044, Italy}
\newcommand{\LNLINFN}{INFN Laboratori Nazionali di Legnaro, Legnaro (Padova) 35020, Italy}
\newcommand{\Lodz}{Institute of Applied Radiation Chemistry, Lodz University of Technology, 93-590 Lodz, Poland}
\newcommand{\LPNHE}{LPNHE, CNRS/IN2P3, Sorbonne Universit\'e, Universit\'e Paris Diderot, Paris 75252, France}
\newcommand{\Mainz}{Institut f\"ur Kernphysik, Johannes Gutenberg-Universit\"at Mainz, D-55099 Mainz, Germany}
\newcommand{\Manchester}{Department of Physics and Astronomy, The University of Manchester, Manchester M13 9PL, UK}
\newcommand{\MEPhI}{National Research Nuclear University MEPhI, Moscow 115409, Russia}
\newcommand{\MendeleevUniverisity}{Mendeleev University of Chemical Technology, Moscow 125047, Russia}

\newcommand{\MIINFN}{INFN Milano, Milano 20133, Italy}
\newcommand{\MIPoliICA}{Civil and Environmental Engineering Department, Politecnico di Milano, Milano 20133, Italy}
\newcommand{\MIPoliCHE}{Chemistry, Materials and Chemical Engineering Department ``G.~Natta", Politecnico di Milano, Milano 20133, Italy}

\newcommand{\MIUni}{Physics Department, Universit\`a degli Studi di Milano, Milano 20133, Italy}
\newcommand{\MSU}{Skobeltsyn Institute of Nuclear Physics, Lomonosov Moscow State University, Moscow 119234, Russia}
\newcommand{\NAINFN}{INFN Napoli, Napoli 80126, Italy}
\newcommand{\NAUniPHY}{Physics Department, Universit\`a degli Studi ``Federico II'' di Napoli, Napoli 80126, Italy}
\newcommand{\NAUniCHE}{Chemical, Materials, and Industrial Production Engineering Department, Universit\`a degli Studi ``Federico II'' di Napoli, Napoli 80126, Italy}
\newcommand{\NAUniPHARM}{Pharmacy Department, Universit\`a degli Studi ``Federico II'' di Napoli, Napoli 80131, Italy}

\newcommand{\NSU}{Novosibirsk State University, Novosibirsk 630090, Russia}
\newcommand{\OACINAF}{INAF Osservatorio Astronomico di Capodimonte, 80131 Napoli, Italy}
\newcommand{\Petersburg}{Saint Petersburg Nuclear Physics Institute, Gatchina 188350, Russia}

\newcommand{\PIINFN}{INFN Pisa, Pisa 56127, Italy}
\newcommand{\PIUniPHY}{Physics Department, Universit\`a degli Studi di Pisa, Pisa 56127, Italy}
\newcommand{\PNNLaddress}{Pacific Northwest National Laboratory, Richland, WA 99352, USA}
\newcommand{\Princeton}{Physics Department, Princeton University, Princeton, NJ 08544, USA}
\newcommand{\Queens}{Department of Physics, Engineering Physics and Astronomy, Queen's University, Kingston, ON K7L 3N6, Canada}
\newcommand{\RHUL}{Department of Physics, Royal Holloway University of London, Egham TW20 0EX, UK}
\newcommand{\RMTreINFN}{INFN Roma Tre, Roma 00146, Italy}
\newcommand{\RMTreUni}{Mathematics and Physics Department, Universit\`a degli Studi Roma Tre, Roma 00146, Italy}
\newcommand{\RMUnoINFN}{INFN Sezione di Roma, Roma 00185, Italy}
\newcommand{\RMUnoUni}{Physics Department, Sapienza Universit\`a di Roma, Roma 00185, Italy}
\newcommand{\SAINFN}{INFN Salerno, Salerno 84084, Italy}
\newcommand{\SAUni}{Physics Department, Universit\`a degli Studi di Salerno, Salerno 84084, Italy}
\newcommand{\SNOLABaddress}{SNOLAB, Lively, ON P3Y 1N2, Canada}

\newcommand{\Temple}{Physics Department, Temple University, Philadelphia, PA 19122, USA}
\newcommand{\TNFBK}{Fondazione Bruno Kessler, Povo 38123, Italy}
\newcommand{\TNTIFPA}{Trento Institute for Fundamental Physics and Applications, Povo 38123, Italy}

\newcommand{\TOINFN}{INFN Torino, Torino 10125, Italy}
\newcommand{\TOPoli}{Department of Electronics and Communications, Politecnico di Torino, Torino 10129, Italy}

\newcommand{\TRIUMFaddress}{TRIUMF, 4004 Wesbrook Mall, Vancouver, BC V6T 2A3, Canada}

\newcommand{\UB}{Universiatat de Barcelona, Barcelona E-08028, Catalonia, Spain} 
\newcommand{\UCDavis}{Department of Physics, University of California, Davis, CA 95616, USA}

\newcommand{\UCLA}{Physics and Astronomy Department, University of California, Los Angeles, CA 90095, USA}
\newcommand{\UMass}{Amherst Center for Fundamental Interactions and Physics Department, University of Massachusetts, Amherst, MA 01003, USA}

\newcommand{\USP}{Instituto de F\'isica, Universidade de S\~ao Paulo, S\~ao Paulo 05508-090, Brazil}
\newcommand{\VTech}{Virginia Tech, Blacksburg, VA 24061, USA}
\newcommand{\WilliamsCollege}{Williams College, Physics Department, Williamstown, MA 01267 USA}
\newcommand{\Zaragoza}{Centro de Astropart\'iculas y F\'isica de Altas Energ\'ias, Universidad de Zaragoza, Zaragoza 50009, Spain}
\newcommand{\ZaragozaARAID}{Fundaci\'on ARAID, Universidad de Zaragoza, Zaragoza 50009, Spain}
\begin{document}

%\author{The DarkSide-20k Collaboration:}
%\input{AU-20k-JINST.tex}
\collaboration{The DarkSide-20k Collaboration}
\author[1]{P.~Agnes}
\author[2,3]{S.~Albergo}
\author[4]{I.~F.~M.~Albuquerque}
\author[5]{T.~Alexander}
\author[6,7]{A.~Alici}
\author[8]{A.~K.~Alton}
\author[9]{P.~Amaudruz}
\author[6,7]{S.~Arcelli}
\author[4]{M.~Ave}
\author[10]{ I.~Ch.~Avetissov}
% ...

\affiliation[1]{\Houston}
\affiliation[2]{\CTINFN}
\affiliation[3]{\CTUNI}
\affiliation[4]{\USP}
\affiliation[5]{\PNNLaddress}
\affiliation[6]{\BOUniPHY}
\affiliation[7]{\BOINFN}
\affiliation[8]{\Augustana}
\affiliation[9]{\TRIUMFaddress}
\affiliation[10]{\MendeleevUniverisity}
% ...
\emailAdd{ds-ed@lngs.infn.it}

\title{Sensitivity of  future  liquid argon dark matter search experiments to core-collapse supernova neutrinos}
%\author{DarkSide Collaboration }
%\linenumbers

%\author[a,1]{S.~Omeone\note{Corresponding author.}}
%\author[a]{S.~Omeone}
%\author[b]{and S.~Omebodyelse}
%\affiliation[a]{Some institute}
%\affiliation[b]{Some university}
%\emailAdd{contact@darkside.it}

\abstract{
Future liquid-argon \DSk\ and \Argo\   detectors, designed for direct dark matter search,   will be sensitive also to core-collapse supernova neutrinos, via coherent elastic neutrino-nucleus scattering. This  interaction channel is flavor-insensitive with a  high-cross section, enabling for a high-statistics neutrino detection with target masses of $\sim$50~t and $\sim$360~t for \DSk\ and \Argo\, respectively. 

Thanks to the low-energy threshold of $\sim$0.5~keV$_{nr}$ achievable by exploiting the ionization channel, \DSk\  and \Argo\ have  the potential to discover  supernova bursts  throughout our galaxy and up to the Small Magellanic Cloud, respectively, assuming a 11-M$_{\odot}$ progenitor star.  We report also on the  sensitivity to the neutronization burst, whose electron neutrino flux is suppressed by oscillations when detected via charged current and elastic scattering. Finally, the accuracies in the reconstruction of the average and total neutrino energy in the different phases of the supernova burst, as well as  its time profile,  are also discussed, taking into account the expected background and the detector response.

}

\keywords{supernova neutrinos, core-collapse supernovae, dark matter detectors, coherent elastic neutrino nucleus scattering}

\arxivnumber{}

\maketitle
\flushbottom

% !TEX root = main.tex

\section{Introduction} 

Core-collapse supernovae (SNe) are violent explosions of very massive stars at the end of their lives, triggered by the gravitational collapse of the stellar cores \cite{burrows1995nature}. The characteristic energy emitted by a core-collapse SN   is $\sim$10$^{53}$~erg, which corresponds to the gravitational binding energy of a 1.4 M$_{\odot}$ core  that collapses into a neutron star. 99\% of this energy is emitted as neutrinos,  $\sim$1\%  goes into the kinetic energy associated with the external layers of the progenitor that are ejected at $\sim$10,000 km/s, and only 0.01\% is radiated at UV, optical and near-infrared wavelengths. Therefore neutrinos are the ideal ``messengers'' for investigating the final stages of stellar evolution, even when the SN is not accessible to optical and radio telescopes \cite{Kamiokande_SN1987A,IMB_SN1987A,Baksan_SN1987A,janka2007theory}. Observations of a neutrino burst from SN 1987A have suggested that the formation of a neutron star might have occurred inside the SN remnant, nevertheless, this fact has been never unambiguously confirmed. SN can play also a key role in the neutrino physics, by providing constraints to the neutrino absolute mass and mass ordering \cite{Horiuchi_Kneller_2018,Mirizzi:2015eza}.

To date, the only  SN observed through neutrinos is the SN 1987A, with a total of 25 events detected by Kamiokande-2 \cite{Kamiokande_SN1987A}, IMB \cite{IMB_SN1987A} and Baksan \cite{Baksan_SN1987A}. Since then, core-collapse SN simulations have made several breakthroughs, providing detailed understanding of the neutronization, accretion, and cooling phases \cite{janka2007theory,Mirizzi:2015eza}. The next detection of galactic SN neutrinos will provide key elements to our comprehension of the mechanisms governing the core-collapse and also on fundamental questions in  neutrino physics.

This paper presents a sensitivity study on SN neutrino detection with the Global Argon Dark Matter Collaboration (GADMC) liquid-argon (LAr) experiments, \DSk\ and \Argo. \DSk\ is a dual-phase time-projection-chamber (TPC) of about \SI{50}{\tonne} mass \cite{DS20k_2018}, designed for dark matter detection, but also sensitive to  low energy nuclear recoils (NR) induced by SN neutrinos via  coherent elastic neutrino-nucleus scattering (\CEnNS) \cite{akimov2017observation}, in construction at Laboratori Nazionali del Gran Sasso LNGS), Italy. The GADMC is also considering a future  single-phase or dual-phase multi-hundred tonne detector, called \Argo, with SNOLAB, Canada, as the preferred location. For this work we assume  that \Argo\ is a dual-phase TPC  with a target mass of \SI{370}{\tonne}.  

%\red{\sout{Neutrino detection via  \CEnNS\ offers a unique opportunity, since it is equally sensitive to all neutrino flavours and therefore allows to measure the unoscillated SN neutrino flux. Current and future giant (kilotons and megatons target mass) detectors, in fact, are mostly sensitive to  electron neutrinos: water-Cherenkov detectors like Super-Kamiokande \cite{Ikeda:2007sa}, Hyper-Kamiokande \cite{DiLodovico:2015kta}, IceCube \cite{Aartsen:2013nla}, and KM3NeT \cite{Molla:2019nns} rely on the  inverse beta decay (IBD) channel;  the DUNE \cite{Abi:2018dnh} LAr TPC  will exploit the $\nu_e \, ^{40}\text{Ar} \rightarrow ^{40}\text{K}^* \, e^-$ charge current interaction;  scintillator detectors like JUNO \cite{An:2015jdp}  will look at   IBD  and  elastic scattering channels. }}

Neutrino detection via \CEnNS\ offers a unique and synergistic opportunity to explore the neutrino signal from a SN and understand the neutrino oscillation physics involved, since it is equally sensitive to all neutrino flavours and therefore allows to measure the unoscillated SN neutrino flux.  Current and future giant (kilotons and megatons target mass) detectors, in fact, are mostly sensitive to electron neutrinos and electron anti-neutrinos:  water-Cherenkov and scintillator detectors, like Super-Kamiokande \cite{Ikeda:2007sa}, Hyper-Kamiokande \cite{DiLodovico:2015kta}, IceCube \cite{Aartsen:2013nla},  KM3NeT \cite{Molla:2019nns}, and JUNO \cite{An:2015jdp},  rely  on the electron antineutrino detection via inverse beta decay (IBD) and are  sensitive to  electron-neutrinos via elastic scattering, whereas the DUNE \cite{Abi:2018dnh} LAr TPC will exploit the electron neutrino charge current interaction  ($\nu_e \, ^{40}\text{Ar} \rightarrow ^{40}\text{K}^* \, e^-$).  The truly flavour-blind measurement of the neutrino signal  via \CEnNS\  yields the normalization of the total flux of SN neutrinos, and could potentially provide a measurement of the neutrino mass hierarchy in combination with the other  experiments.

An additional advantage of the \CEnNS\ channel  is the high cross-section,  roughly 50 times larger  than  that of charge current interaction \cite{Scholberg_supernova_2012} at 10 MeV, which compensates for the relatively small target masses of \DSk\ and \Argo, and which allows for  high-statistics detections.

The sensitivity to SN neutrino detection via \CEnNS\ process has been thoroughly investigated for future liquid xenon  dark matter detectors like XENONnT, DARWIN, and LZ  \cite{PhysRevD.94.103009, Khaitan:2018wnf}.  Although the lower LAr density imposes larger TPC volumes with respect to liquid xenon experiments, and hence a slightly worse time resolution due to the longer drift time, LAr experiments can provide a better energy resolution. The  lighter argon nucleus and the smaller energy quenching effect, in fact,  as demonstrated in this work,   provide  higher sensitivity to SN burst parameters that can be inferred from the nuclear recoil energy spectrum induced by SN neutrinos. In addition, the lower energy threshold allows for  larger statistics, compensating for the lower cross-section with respect to liquid xenon targets.

 In this work, we provide an extensive study for argon detectors, assuming a  background level derived from the most recent contamination measurements from material screenings. After a detailed description of the expected signal (section \ref{sec:CCSN}), of the detector response (section \ref{sec:signal}) and of the expected background (section \ref{sec:background}), we discuss the  \DSk\ and \Argo\ discovery potential to SN burst in section \ref{sec:analysis}. We will also report on the sensitivity to the neutronization burst and to the mean and integrated neutrino energies from the SN accretion and cooling phases.

% !TEX root = main.tex

\section{Core-collapse supernovae and neutrinos} 
\label{sec:CCSN}

A very massive star can undergo core-collapse when, at the end of its life, the iron core of the progenitor star, grows to roughly the Chandrasekhar mass, and nuclear fusion can no longer balance  the inward push from the force of gravity. In this regime,
neutrinos are mostly produced by electron captures on heavy nuclei and  leave the core unimpeded. After a few milliseconds,  the neutrino mean free path becomes comparable to the core radius and neutrinos remain trapped in ultra-dense matter. \cite{janka2001conditions}.
Despite the trapping,   neutrinos  around the newly formed neutrinosphere can still escape.

When compression of matter reaches a critical density,  the core rebounds. The violent rebound of the matter produces a pressure wave propagating outwards, which eventually steepens into a shock wave, and neutrino emission again increases rapidly, producing the so-called neutronization burst, lasting about 30 ms. The shock, in fact,  is so powerful that it dissociates nuclei into free nucleons all along its way to the edge of the core. Free protons quickly interact with the energetic electrons, resulting in neutrons and electron neutrinos. 
 \cite{Bruenn_etal_2013,bruenn2009mechanisms}.

Neutrinos are the only messengers that can bring us direct information about the neutronization phase.  During their propagation through the stellar mantle to Earth, neutrinos oscillate,  with a flavor conversion amplified by the Mikheev-Smirnov-Wolfenstein (MSW) effect \cite{MSW_resonance_1986}, in agreement with the matter density profile crossed.  Additional phenomena, such as  matter turbulence, fluctuations in stellar matter density, and neutrino-neutrino interactions, can lead to alteration of the MSW effect, and hence of neutrino flavor conversion. As a net effect, the survival probability at the Earth of $\nu_e$'s, produced in the neutronization phase, is expected to be $\sim$2\% ($\sim$30\%) assuming the normal (inverted) mass ordering \cite{Horiuchi_Kneller_2018,duan2007neutrino,lunardini2003probing}. 

%\red{\sout{This electron flavor suppression, together with the low statistics, did not allow the neutronization burst  to be observed in SN1987A. Even future experiments, primarily  relying on charge current interactions, will be significantly limited in their sensitivity to the  neutronization  burst. In contrast, neutrino flavor conversion does not affect the results reported in this work, as \CEnNS\ is flavor insensitive, and therefore GADMC LAr TPCs will be able to detect the entire SN neutrino flux.   Furthermore, it is interesting to observe how the comparison of the interaction rates measured by these TPCs with the future charge current measurements, mentioned in the previous section, will allow  improving constraints on the neutrino mass ordering. }}

Since the neutronization burst is dominated by neutrinos, not  accessible via  IBD, the signal observed from SN1987A did not allow for the neutronization burst to be directly observed. Furthermore, even future experiments that  rely on IBD and charge current interactions but also  on elastic scattering, will be significantly limited in their sensitivity to the neutronization burst due to the flavor suppression, especially in the case of a normal mass hierarchy.  In contrast, the neutrino flavor conversion does not affect the results reported in this work, as \CEnNS\  is flavor insensitive, and therefore GADMC LAr TPCs will be able to detect the entire SN neutrino flux.

After the neutronization, the shock wave may stall losing energy in the dissociation of the nuclei, thus being unable to overcome the ram pressure of the material falling into the shock. Neutrinos can revitalize the shock, depositing energy into the envelope. This critical stage, named accretion phase,  lasts a few hundred milliseconds and can lead either to the star explosion or to its collapse, and thus to the formation of a black hole. Multi-dimensional simulations suggest a standing accretion shock instability (SASI) \cite{Tamborra_etal_2013}, where the shock front oscillates inward and outward, periodically, leading to a $\mathcal{O}(10-100)$~Hz modulation of the neutrino luminosity. Although this effect can potentially be observed with GADMC TPCs, thanks to the time resolution in the millisecond range, the present work is based on 1D simulations, and therefore sensitivity to SASI will not be discussed. 

\begin{figure}[]
  \centering 
    \includegraphics[width=0.8\textwidth]{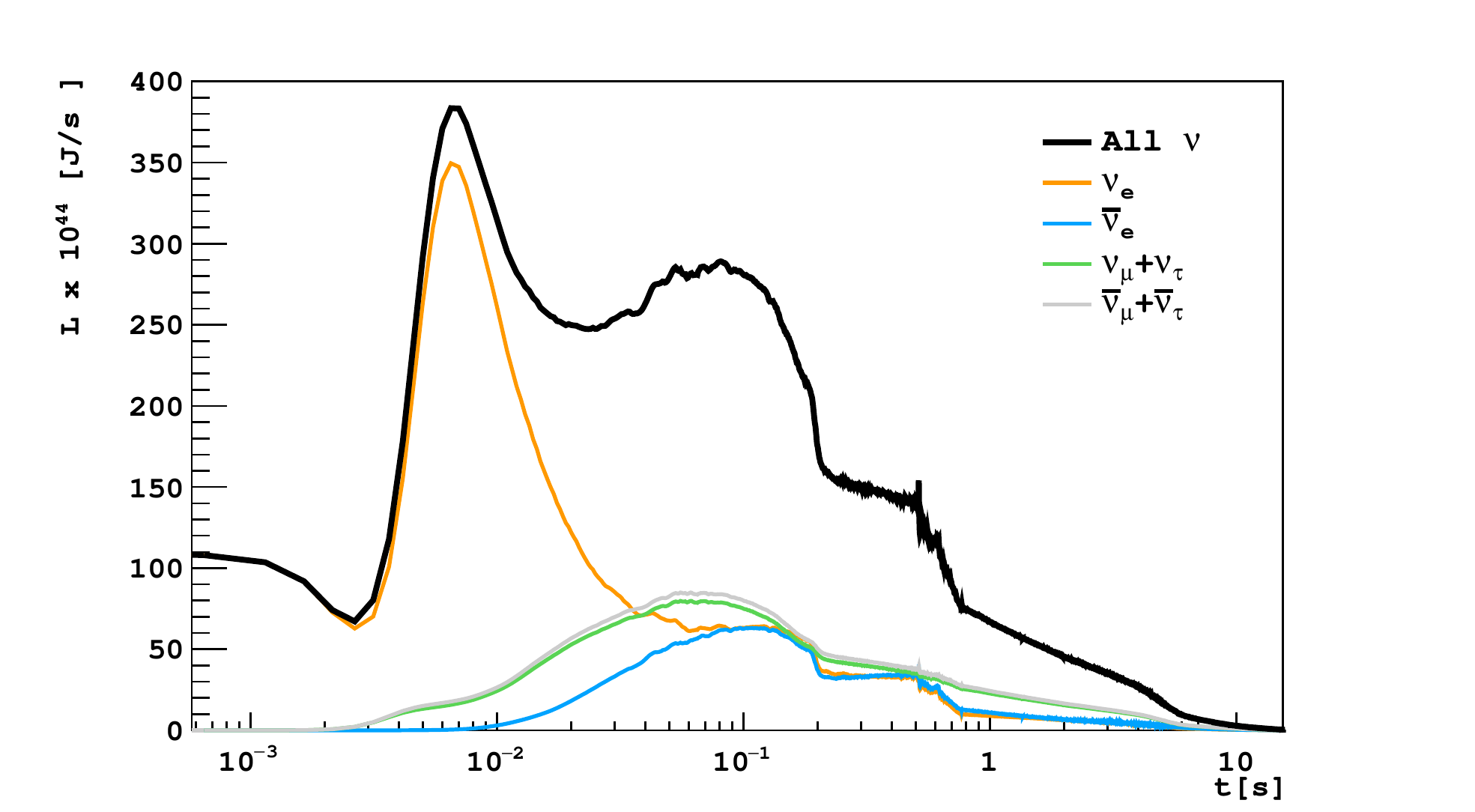}
    \includegraphics[width=0.8\textwidth]{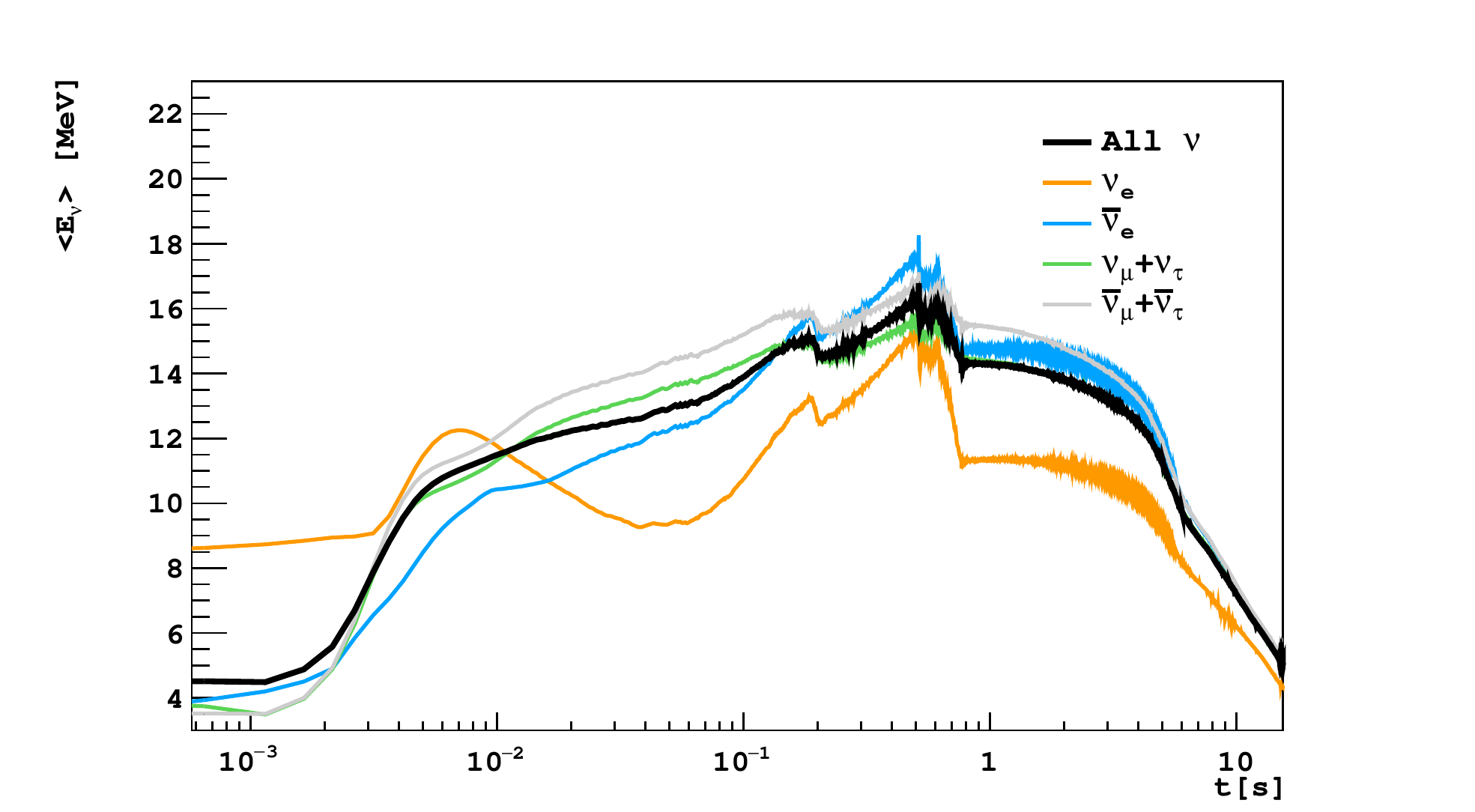} 
    \includegraphics[width=0.8\textwidth]{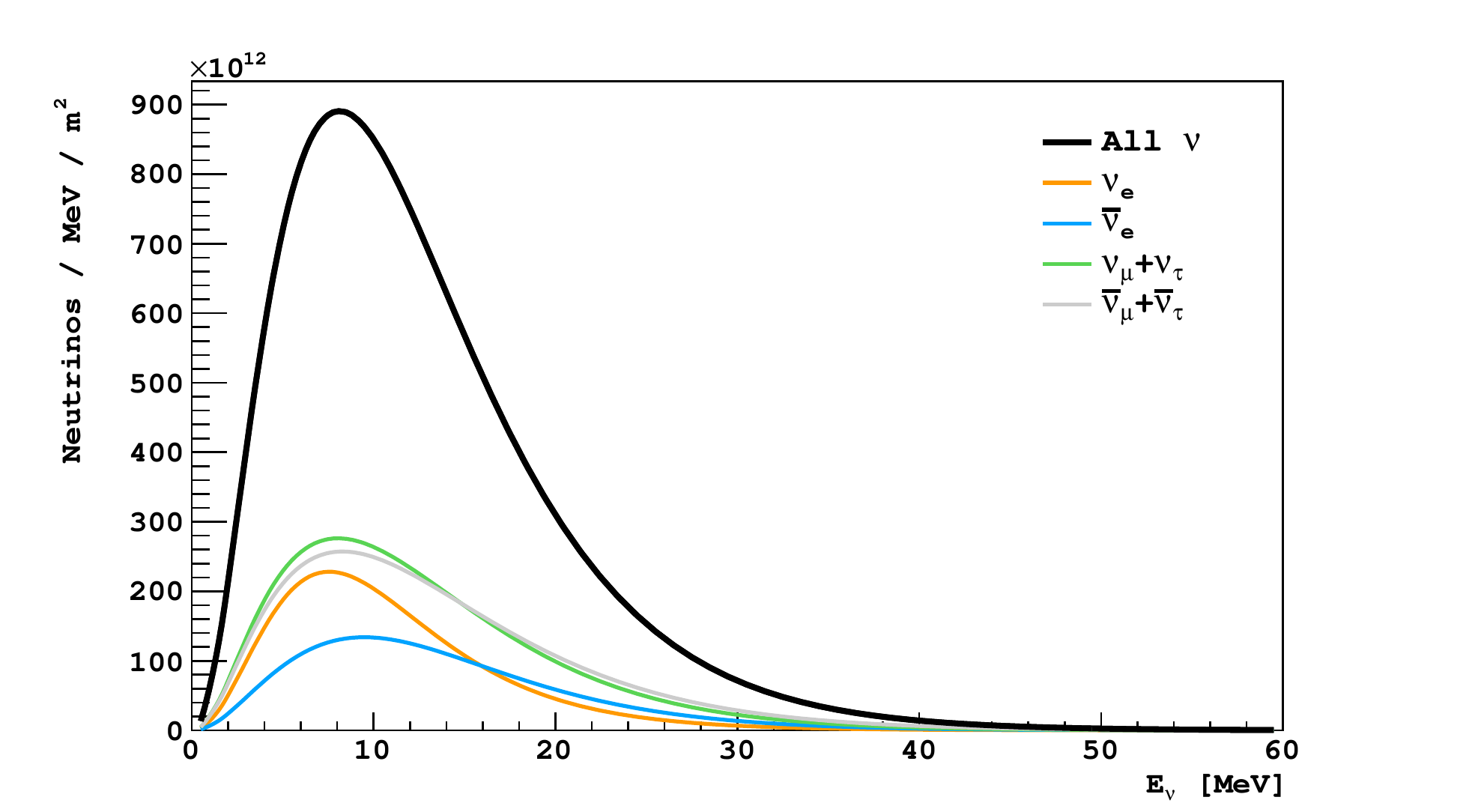}
 
    \caption{ Time evolutions of neutrino luminosity (top) and mean energy (middle) and energy spectrum (bottom) from a core-collapse \LargeSolarMass\ SN  for the different neutrino species, using   Garching group 1-d simulations  \cite{Bruenn_etal_2013}. 
    }
    
  \label{fig:NeutrinoLuminosity}
\end{figure} 

The explosion of the SN  blows off almost all the matter in the stellar mantle and leaves the hot proto-neutron star. The third phase, the cooling of the neutron star by neutrino emission, lasts about \SI{10}{\second} \cite{Bruenn_etal_2013}. The neutrino mean energy \NeutrinoMeanEnergy\ drops from \SI{15}{\MeV} to  \SI{5}{\MeV}  in about \SI{10}{\second}, while the neutrino luminosity  decreases roughly according to the law of black body radiation \cite{chiu1960neutrino}.

The luminosity and mean energy time evolutions and the energy spectrum are shown in \reffig{NeutrinoLuminosity} from 1-d  hydrodynamical spherically symmetric core-collapse SN simulations by the Garching group \cite{Mirizzi:2015eza, Huedepohl, Garching},  using the LS220 equation of state, for a  progenitor star mass  of \LargeSolarMass. This is the reference model adopted in this work, and we will report results also for a progenitor star mass  of \SmallSolarMass.

% !TEX root = main.tex

\section{Supernova neutrino signal and detector response} 
\label{sec:signal}

The \CEnNS\  differential cross-section as a function of  neutrino energy, $E_\nu$, and recoil energy, $E_r$,  is given by 
\begin{equation}
\mathrm{d}\sigma(E_\nu,E_r) = \frac{{G^2_F}}{4\pi} {Q^2_W} m\left(1-\frac{m E_r}{2{E^2_\nu}}\right) F^2(q)\, \mathrm{d}E_r,
\label{eq:cenns}
\end{equation}
where $G_F$ is the Fermi coupling constant, $Q_W$ the weak charge of argon nucleus,  and $m$ the argon nucleus mass. $F(q)$ is the Helm form factor, parametrized with the Lewin-Smith approach \cite{Lewin:1995rx}, as a function of the momentum transfer $q=\sqrt{2m E_r}$. 

\begin{figure}[htb!]
  \centering 
    \includegraphics[width=0.8\textwidth]{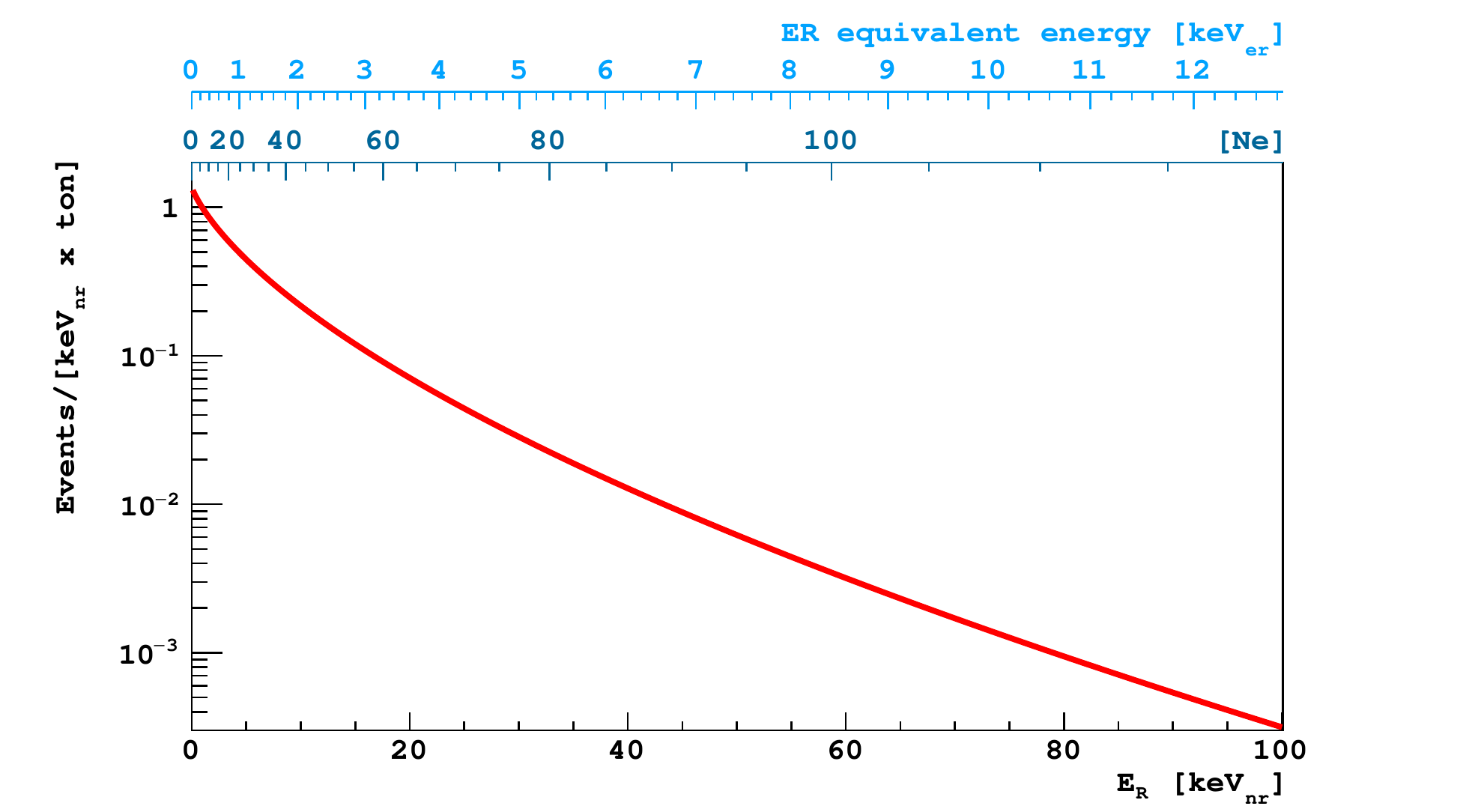}
    \caption{Nuclear recoil energy spectrum from   neutrino interactions in LAr via \CEnNS\  from a core-collapse \LargeSolarMass\ supernova at  \DSkSupernovaBaseline.}
  \label{fig:RecoilSpectrum}
\end{figure}

The  nuclear recoil (NR) energy spectrum induced by SN neutrinos, shown in  \reffig{RecoilSpectrum},  results from the convolution of  the  neutrino flux with the differential neutrino cross-section from eq. \ref{eq:cenns}. The   window of observation is  $<$100~keV$_{nr}$, with $\sim$70\% ($\sim$50\%) of events with energy $<$10~keV$_{nr}$ ($<$5~keV$_{nr}$ ). The low energy detection threshold, therefore, plays a crucial role in the final sensitivity.

The detection mechanism of interacting particles in dual-phase GADMC LAr TPCs relies on a prompt light pulse (S1) induced by scintillation, followed by a delayed pulse  (S2) associated to  ionization  electrons. These, in fact,  are drifted vertically upwards by the drift field, and extracted, by the so-called extraction field, in a thin layer of gas, where they induce a secondary light signal by electroluminescence. 

The detection efficiency of S1 photons is estimated in \DSk\  at 19\% through Monte Carlo simulations. Therefore, a detection strategy based on  S1 photons trigger, highly inefficient for NRs in the keV$_{nr}$ range, would strongly affect the  sensitivity to SN neutrinos.   Disregarding S1, S2 guarantees an amplification factor by more than 20 ($\sim$23 photoelectrons per electron extracted in the gaseous phase in \DSf~\cite{LowMass}), allowing the detection of NRs with a threshold of a few hundreds of eV$_{nr}$. This approach was  successfully applied by \DSf\ in  setting the world's best limit on WIMP dark matter particles in the 2-6 GeV/c$^2$ mass range~\cite{LowMass}, with a $\sim$0.6~keV$_{nr}$ threshold. In the same work, \DSf\  demonstrated  a detection efficiency at 100\% level for NR  deposits with an energy of 0.46 keV$_{nr}$, allowing the detection of about 86\% of NRs induced by SN neutrinos. 

The dual-phase LAr TPC response to NRs, in the S2 channel,  differs from the one to electronic recoils  (ERs), which account for almost all of the background. This is due to the differences between ER and NR excitons to ionization electrons ratio, as well as to the recombination process, which  produces excited argon dimers and depletes  the ionization channel.  In addition, the largest fraction of energy deposited by NRs is  neither converted into scintillation nor ionization, resulting in a quenching effect much stronger than that observable for ERs \cite{Agnes:2018mvl}. 

The NR energy scale in the S2 observable was determined with $^{241}$Am--$^{9}$Be  and $^{241}$Am--$^{13}$C neutron sources~\cite{LowMass} deployed outside the \DSf\ cryostat, and from  neutron-beam scattering data from the SCENE \cite{cao2015measurement} and ARIS~\cite{Agnes:2018mvl} experiments. The S2 ER energy scale is obtained from \DSf\ data by fitting the Thomas-Imel model \cite{PhysRevA.36.614} to the mean S2 measured for the 2.82~keV K-shell and 0.27~keV L-shell lines from the electron capture of the cosmogenic $^{37}$Ar \cite{Agnes:2018oej}.  At the nominal drift field of 200 V/cm at which GADMC TPCs operate, and using S2 as energy variable,   the  ER energy corresponding to 100~keV$_{nr}$ is  about 13 keV$_{er}$, as shown in \reffig{RecoilSpectrum}.

The energy resolution model adopted in this work accounts for the LAr intrinsic fluctuations of the ionization and electron-ion recombination processes, and for the statistics governing the  emission and detection of photons induced by electrons in the gas pocket. Intrinsic processes fluctuate with respect to the binomial probability defined as the ratio of the number of free ionization electrons and the number of all quanta produced by the particle interaction. The latter is obtained by dividing the deposited energy  by the effective  work function in LAr (19.5~eV~\cite{Agnes:2017grb}). The photoelectron statistics is assumed normal, with a photon yield of 23 photoelectrons per ionization electron, in agreement with the \DSf\ measurement. 

The event time resolution is dominated by the electron drift time, which, in absence of a S1 pulse,  induces a delay with respect to the SN neutrino interaction time. The drift velocity in presence of an electric field of  200 V/cm is (0.93 $\pm$ 0.01) mm/$\mu$s, which corresponds to a maximum drift time, T$_{max}$  of $\sim$3.8 ms in \DSk\ (3.5 m height), and of $\sim$5.4 ms in \Argo\ (5.0 m height). As SN neutrino events are uniformly distributed in the TPC, the corresponding standard deviations, calculated as T$_{max}$/$\sqrt{12}$,  are $\sim$1.1~ms and $\sim$1.6~ms, respectively.  

The same response model is applied to energy deposits from the background sources discussed in the next section.

\section{Expected background in GADMC TPCs}\label{sec:background}

The \DSk\ (\Argo) TPC  is an octagonal regular prism with a distance  of 3.5~m (8~m)  between parallel lateral walls, resulting in a total active LAr mass of 49.7~t (371~t). Differently from \DSf, where the TPC is housed in a stainless steel cryostat,  \DSk\ and \Argo\ TPCs will be enclosed in an acrylic envelope, characterized by a larger radio-purity and smaller  mass. This will be possible thanks to the new design, where the TPC is entirely immersed in a LAr bath within a proto-DUNE-like cryostat, serving as active and passive shielding against cosmic rays and environmental radioactivity, respectively. In this new design,  photomultiplier tubes  that detect light in  \DSf\ will be  replaced by silicon photomultipliers (SiPMs), which provide higher quantum efficiency and radiopurity \cite{ESPP,  DS20k_2018}.

The background expected in the energy range of observation for SN neutrinos ($<$100~keV$_{nr}$) can be inferred from the one measured in \DSf.  Above $\sim$1~keV$_{nr}$, this is  dominated  by LAr intrinsic contamination from $^{39}$Ar and $^{85}$Kr $\beta$-decays,  and by radioactivity from the detector materials surrounding the active mass.   

\begin{figure}
   \centering
    \includegraphics[width=0.8\linewidth]{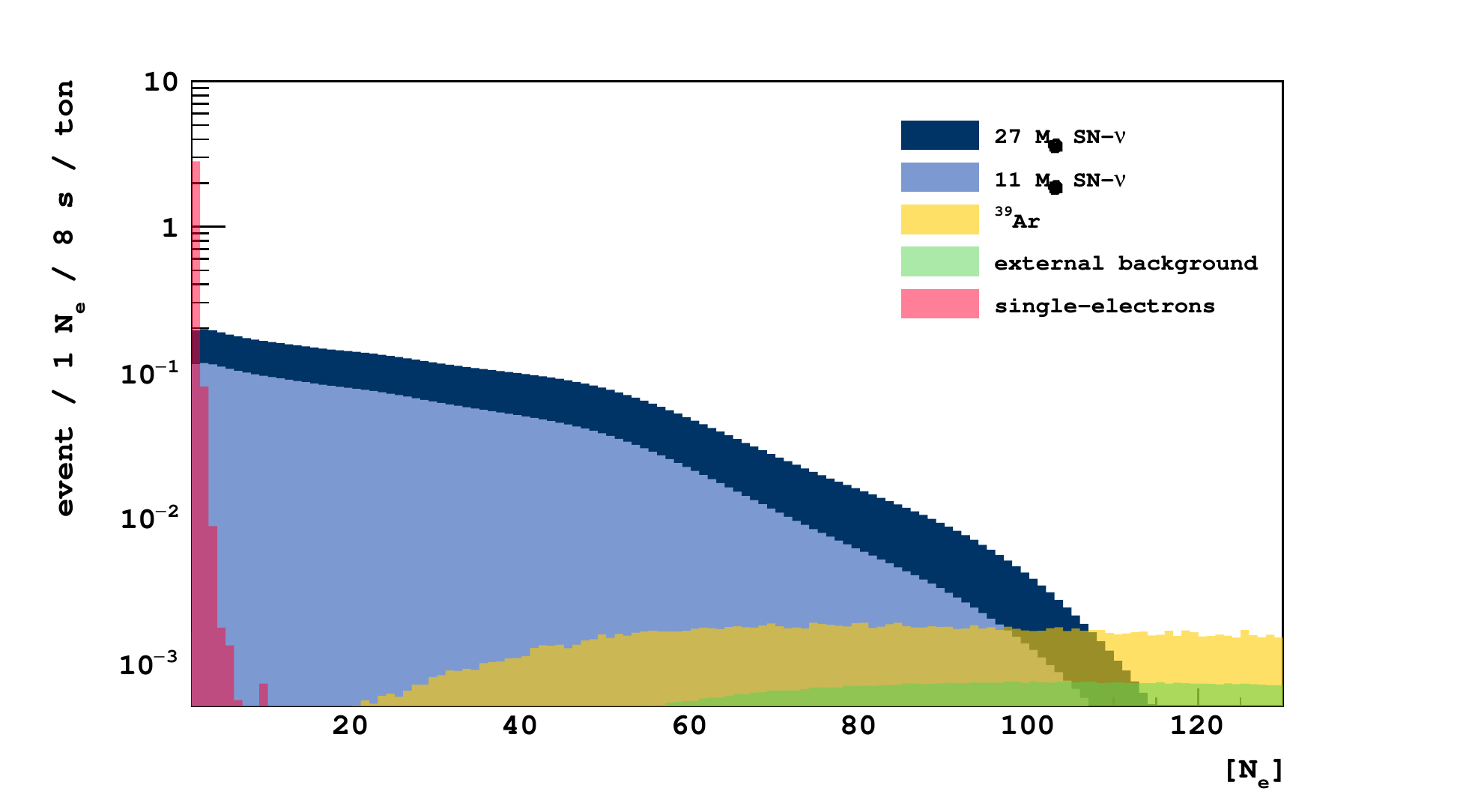}
   \includegraphics[width=0.8\linewidth]{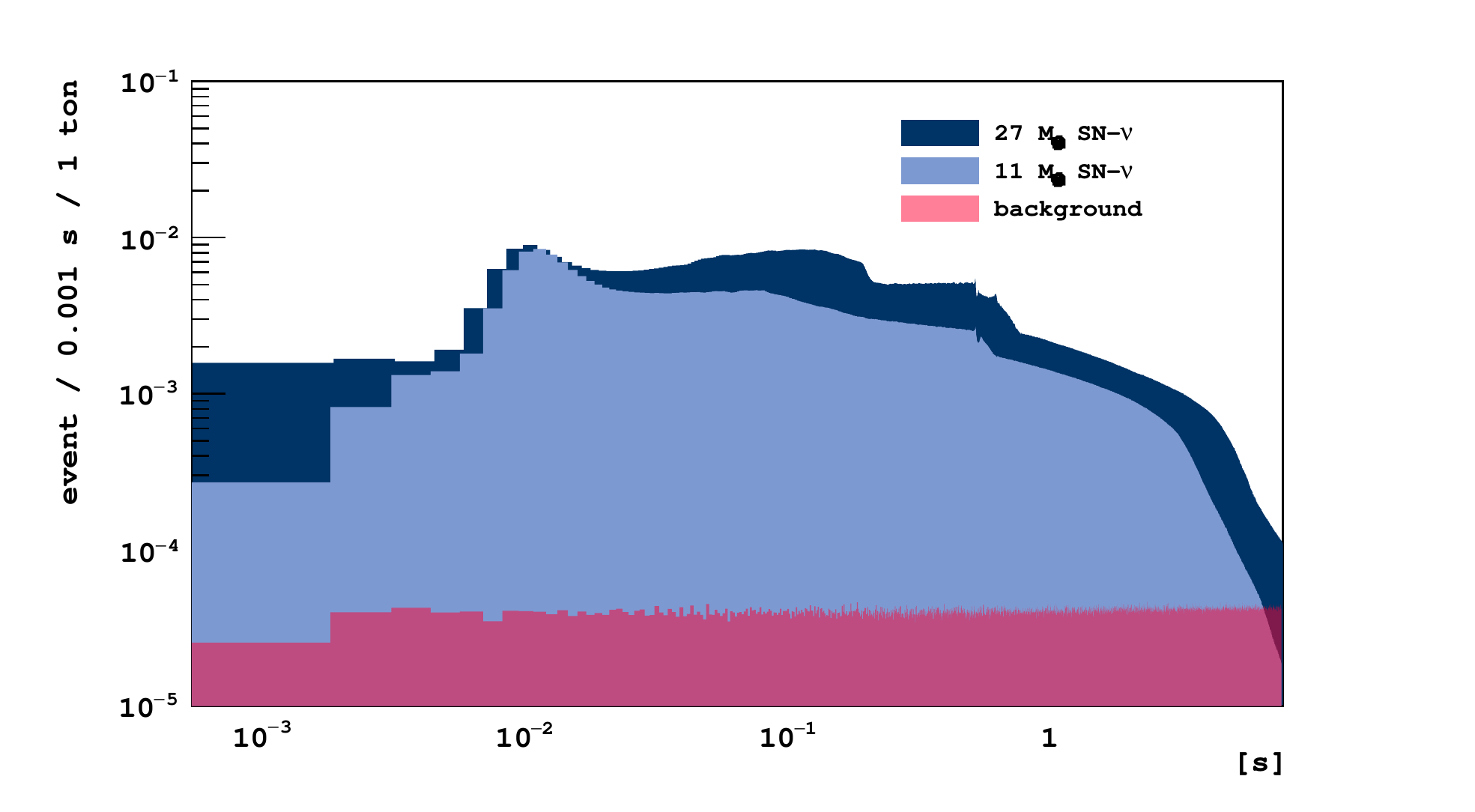}
   \caption{Top. Energy spectrum in number of ionization electrons (\nel) per unit of mass of neutrinos from 11-M$_{\odot}$ and 27-M$_{\odot}$ SNe and background from  single electron events,  \ArThirtyNine\ decays and external background from SiPMs. Bottom. Time evolution of signal and all  background components (external background as expected in \Argo) by selecting events in the [3,100]~\nel\ energy range.   }
    \label{fig:background}
\end{figure}

$^{39}$Ar has a cosmogenic origin, as it is produced by cosmic rays via spallation on $^{40}$Ar. In order to suppress such a background, the LAr  active mass is extracted from deep underground wells  (UAr) in Cortez, Colorado (USA), naturally shielded against cosmic rays. \DSf\ has measured an $^{39}$Ar specific activity  of $\sim$0.7~mBq/kg. In the same campaign,  \kre\ was identified with a specific activity of  $\sim$2~mBq/kg.  The anthropogenic nature of  \kre\ suggests  tiny air contamination in UAr occurred  during the detector filling, possibly at the origin also of the residual $^{39}$Ar activity. This hypothesis, corroborated later by the identification of a leak in the purification phase, suggests an even smaller  $^{39}$Ar intrinsic contamination   in UAr. For both \DSk\ and  \Argo, any residual  \kre\  activity will  be entirely suppressed by  distillation thanks to  \Aria, a 350 m tall distillation column in the phase of installation in the Seruci mine in Sardinia  \cite{DS20k_2018}. In this work, \kre\ contamination is therefore assumed negligible, but we consider the  most conservative hypothesis on $^{39}$Ar specific activity, corresponding to the one measured by \DSf\ in UAr. As shown in \reffig{background} (top) that displays the energy distribution of expected signal and background, the contribution from $^{39}$Ar becomes comparable to the signal from a 10~kpc 11-M$_{\odot}$ SN at $\sim$100 number of ionization electrons, \nel, corresponding to $\sim$8.5~keV$_{er}$.  The total expected rate of $^{39}$Ar events in \DSk\ (\Argo) is 0.5 Hz (4.2~Hz), taking into account that  the fraction of $^{39}$Ar events  with \nel $<$100  is $\sim$1.7\%.

The external background rate is estimated from the  contamination, measured in material screening campaigns (not yet completed at the time of writing), of  radioactive chains ($^{238}$U, $^{235}$U and $^{232}$Th) and individual isotopes ($^{137}$Cs, $^{53}$Mn, $^{40}$K, $^{60}$Co). Each contaminant was simulated with G4DS \cite{Agnes:2017grb}, the DarkSide Monte Carlo package, tracking the radiation from the detector components,  primarily  from the acrylic vessel and  SiPMs. Since SN neutrinos interact only once in LAr,  multiple-scatter events,   identified by the detection of multiple S2 pulses, are efficiently rejected.    The  rate of the residual single-scatter events   in \DSk\ (\Argo) is expected  to be  $\sim$75~Hz ($\sim$320~Hz) in the entire energy range. Narrowing in the region of interest for SN neutrinos, the rate drops to $\sim$0.3~Hz ($\sim$1.3~Hz).

Simulations demonstrate that the mean attenuation length in LAr of single-scatter ER events from the external background, with energy less than $<$8.5~keV$_{er}$,  is $\sim$0.5~cm. The external contamination becomes thus negligible  by rejecting events within 5~cm from the detector walls.  The  event position  is  reconstructed at the centimeter level on the plane orthogonal to the electric field, exploiting the S2 signal and the segmentation of the photodetection modules. The active mass resulting from the volume fiducialization is 47.1~t in \DSk\ and 362.7~t in \Argo. 

The events originating from  the upper and lower planes can be ideally suppressed using the dependence of the ionization electron cloud diffusion on the vertical position,  as discussed in  \cite{Agnes:2018hvf}.  However, since we don't have  an estimate of the rejection efficiency at such low energies, the background from the top and bottom planes is conservatively included in this study. Its residual rate is 0.2~Hz in \DSk\ and 1.1~Hz in \Argo.  

%\red{This approach is an alternative to estimating vertical position through drift time, not applicable in this energy range due to the lack of S1.}
%\red{In this work we therefore consider the residual external background in the fiducial mass to be negligible.}

The sub-keV$_{nr}$ energy region is dominated by a large population of spurious electrons, here named "single-electrons", whose origin is still under investigation. A fraction of these events is related to impurities present in LAr that capture drift electrons and re-emit them with a delay that varies from a few milliseconds to several seconds. A time correlation has been observed in \DSf\ between a fraction  of single-electrons and events with an  large amplitude S2 pulse   preceding  them.  However,  the mechanism behind the majority of single electrons remains unknown. 

In this work, we  assume, for the single-electron background, the spectrum  of single-electrons as measured in \DSf, after subtraction of known internal and external background components \cite{LowMass}, scaling the rate by the target mass ratio between \DSf\  and  \DSk\ or \Argo. The single electron  rate measured in \DSf\ is $\sim$380~mHz/ton, and drops to $\sim$1.8~mHz/ton  by applying a threshold cut at \nel$\geq$3, as shown in \reffig{background} for neutrino signals from 11-M$_{\odot}$ and 27-M$_{\odot}$ SNe. Pile-up of single electrons with physics events are expected with probabilities equal to 6\% and 49\% for \DSk\ and \Argo, respectively. The single electron component in such events can be efficiently identified and removed by applying selection cuts on the spatial distance between the two interactions. 
%The only particles that can produce double scattering in the TPC are, in fact, gammas and neutrons. The former induce  interactions distant at the centimeter level in the energy range of interest for this study, far smaller than the TPC size, and the latter are expected with  a negligible rate.     

\begin{figure}[t]
   \centering
    \includegraphics[width=0.7\linewidth]{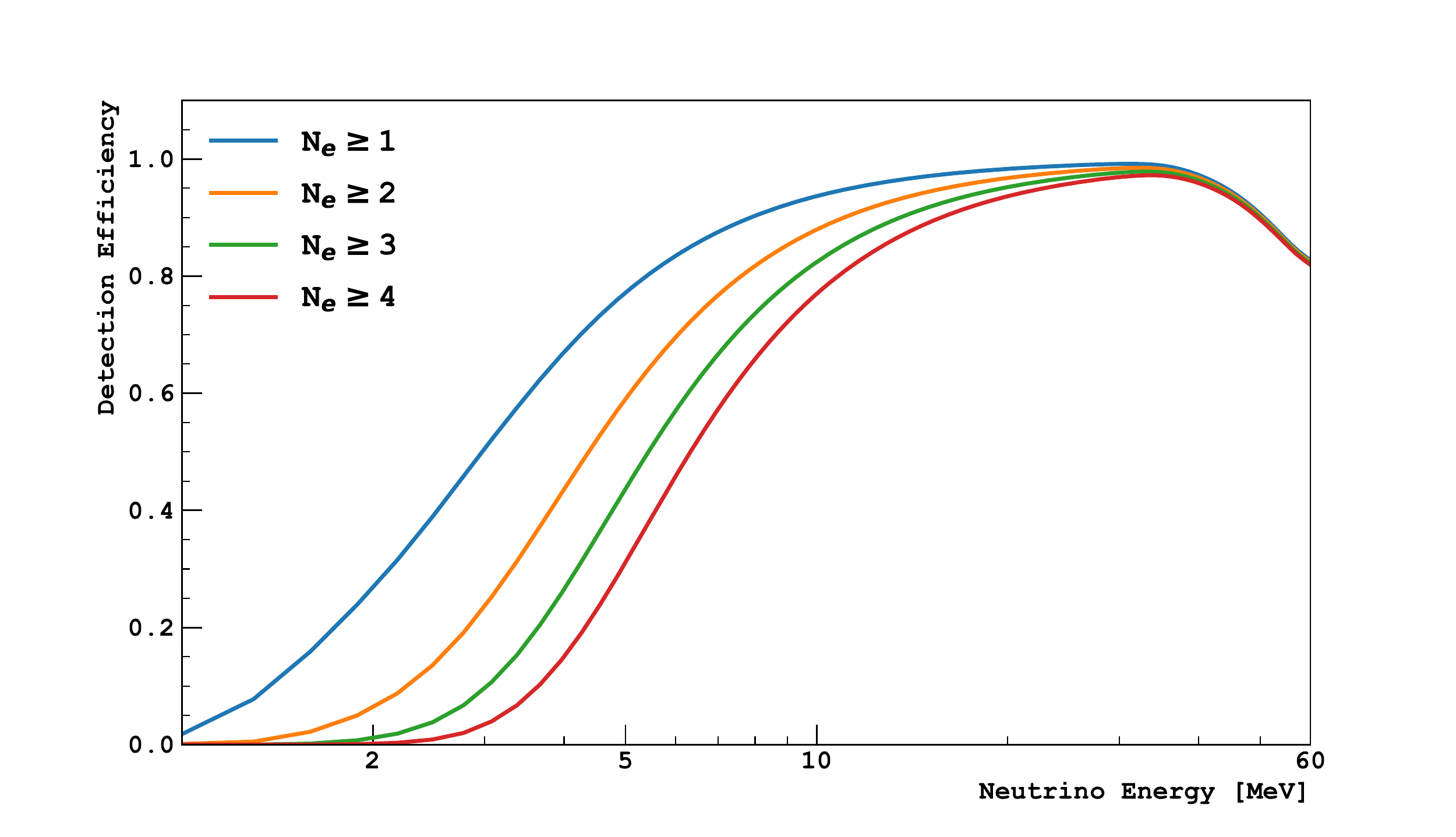}
   \caption{Neutrino detection efficiency via \CEnNS\ as a function of neutrino energy, for different \nel\ thresholds and below 100~\nel.  }
    \label{fig:efficiency}
\end{figure}

The  window of observation is then defined within 8~s from the burst and between 3 and 100~\nel, in order to suppress single-electron background and $^{39}$Ar events, respectively.  The neutrino detection efficiency via \CEnNS\ in the [3, 100]~\nel\ range, shown in \reffig{efficiency}, leads to  expected number of signal events in \DSk\  (\Argo) of 181.4 (1396.6) and 336.5 (2591.6)  from 11-M$_{\odot}$ and 27-M$_{\odot}$ SN burst at 10~kpc, as quoted in table \ref{tab:eventcomparison}. 

The expected overall signal-to-background ratio in the GADMC TPCs for the two SN models is $\sim$24 and $\sim$45, respectively. In particular, as reported in table \ref{tab:signal-to-background},  the signal is about two orders of magnitude larger than the background during the neutronization burst ($<$0.02 s) and the accretion phase ([0.02, 1]~s), while it is about one order of magnitude in the cooling phase ([1, 8]~s), where however the statistic is the largest.

From the same table \ref{tab:signal-to-background}, it can be noticed that the number of events expected from the neutronization burst varies  by only 10\% between 11-M$_{\odot}$ and 27-M$_{\odot}$ SNe,   while those from accretion and cooling phases vary by almost a factor of two. As already suggested in ref. \cite{PhysRevD.94.103009}, the relatively high statistic measurements of the differential energy and time spectra of the SN with Argo, that will be discussed in section \ref{sec:analysis}, can provide a substantial constraint of SN models and  pave the way to the progenitor mass measurement. The sensitivity to the mass is not considered in this work but will be evaluated in the future, once the relationship between  progenitor mass and  fraction of neutrinos emitted during neutronization will be assessed by theory.

\begin{table}
\centering
\setlength{\tabcolsep}{18pt}
\caption{Event statistics expected in \DSk\ and \Argo\ from 11-M$_{\odot}$ and 27-M$_{\odot}$ SNe at 10~kpc and from single-electron and $^{39}$Ar background components, within the  [3, 100]~\nel\ energy window and in 8~s from the beginning of the burst. 
}
\label{tab:eventcomparison}
\begin{tabular}{ lcc  }
\hline 
&  \DSk   & \Argo   \\
\hline 
11-M$_{\odot}$ SN-$\nu$s  &  181.4  & 1396.6  \\
27-M$_{\odot}$ SN-$\nu$s  &  336.5  & 2591.6  \\
$^{39}$Ar  &  4.3  & 33.8 \\
external background  &  1.8  & 8.8 \\
single-electrons  &  0.7  & 5.1  \\
\end{tabular}
\end{table}

%The comparison of neutrino signal from 11- and 27-M$_{\odot}$ SNe and background from single-electrons and $^{39}$Ar, per unit of target mass, is shown in \reffig{background}. 

\begin{comment}
\begin{table}
\centering
\setlength{\tabcolsep}{18pt}
\caption{Number of events per unit of mass expected in GADMC TPCs from 11-M$_{\odot}$ and 27-M$_{\odot}$ SNe at 10~kpc and  signal-to-background ratio, accounting for single-electron and $^{39}$Ar rates, within the  [3, 100]~\nel\ energy window. 
}
\label{tab:signal-to-background}
\begin{tabular}{ lcccc  }
\hline 
&  \multicolumn{2}{c}{11-M$_{\odot}$ SN}   & \multicolumn{2}{c}{27-M$_{\odot}$ SN} \\
SN phase & [1/t] & S/B   &  [1/t] & S/B    \\
\hline 
Burst      & 0.08   & 276 &   0.09 & 315 \\
Accretion  &  1.83  & 137 &   3.30 & 247\\
Cooling    &   1.96 & 20 &  3.76  & 39 \\
\end{tabular}
\end{table}
\end{comment}

\begin{table}
\centering
\setlength{\tabcolsep}{10pt}
\caption{Number of events per unit of mass expected in GADMC TPCs from 11-M$_{\odot}$ and 27-M$_{\odot}$ SNe at 10~kpc and  signal-to-background ratio, accounting for single-electron, external background, and $^{39}$Ar rates, within the  [3, 100]~\nel\ energy window. }
\label{tab:signal-to-background}
\begin{tabular}{ l|ccc|ccc  }
\hline 
&  \multicolumn{3}{c|}{11-M$_{\odot}$ SN}   & \multicolumn{3}{c}{27-M$_{\odot}$ SN} \\
 & SN-$\nu$  &\multicolumn{2}{c|}{S/B}   &  SN-$\nu$ & \multicolumn{2}{c}{S/B}    \\
 SN phase &  [1/t] & DS20k & ARGO  & [1/t] & DS20k & ARGO     \\
\hline 
Burst      & 0.08   & 212 & 231 &   0.09 & 243 & 264 \\
Accretion  &  1.83  & 105 & 114 &   3.30 & 190 & 207\\
Cooling    &   1.96 & 16 &  17 & 3.76  & 30 & 33 \\
\end{tabular}
\end{table}
% !TEX root = main.tex

\section{Sensitivity to supernova neutrinos} 
\label{sec:analysis}

\begin{figure}
   \centering
    \includegraphics[width=0.8\linewidth]{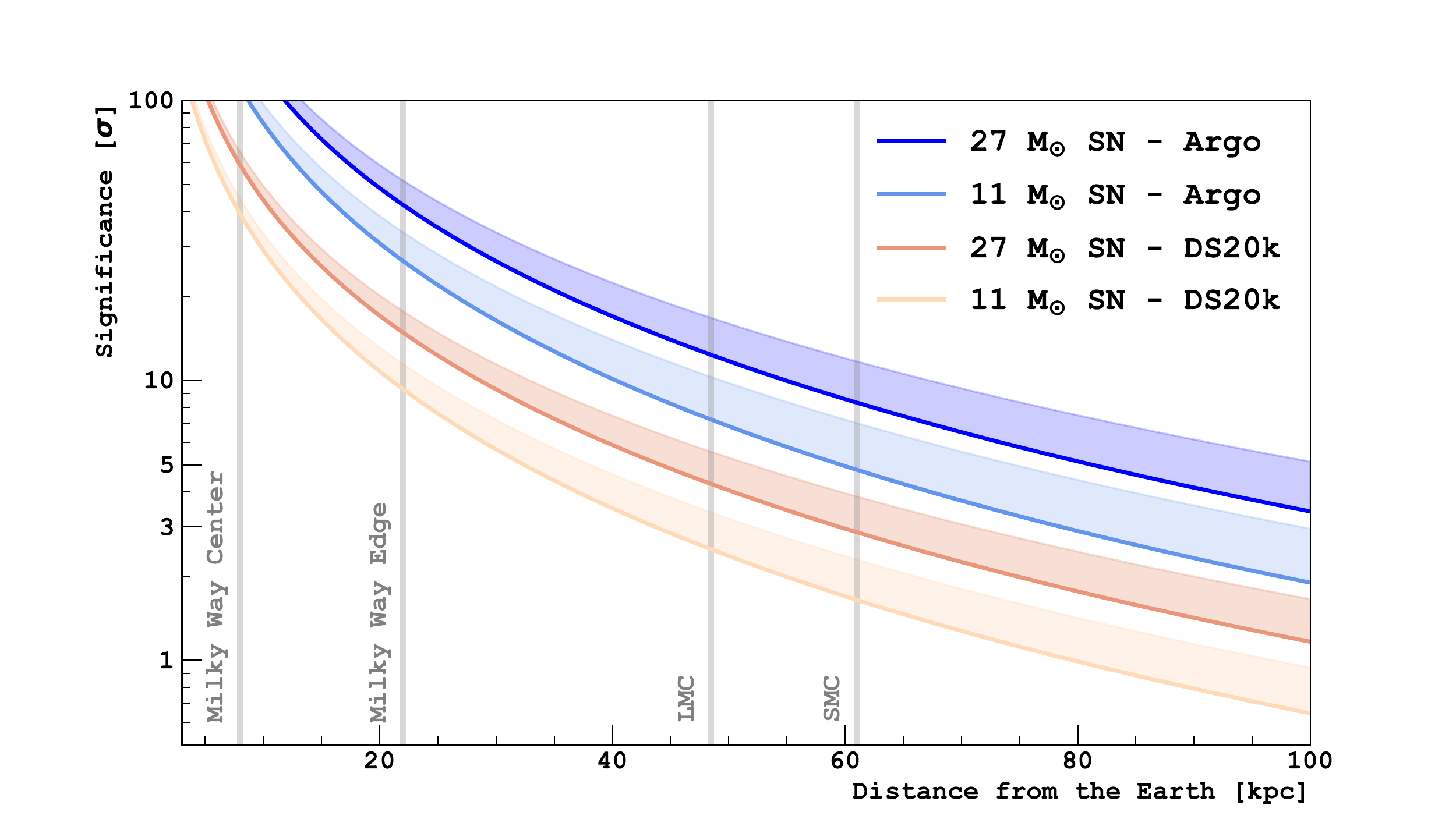}
   \includegraphics[width=0.8\linewidth]{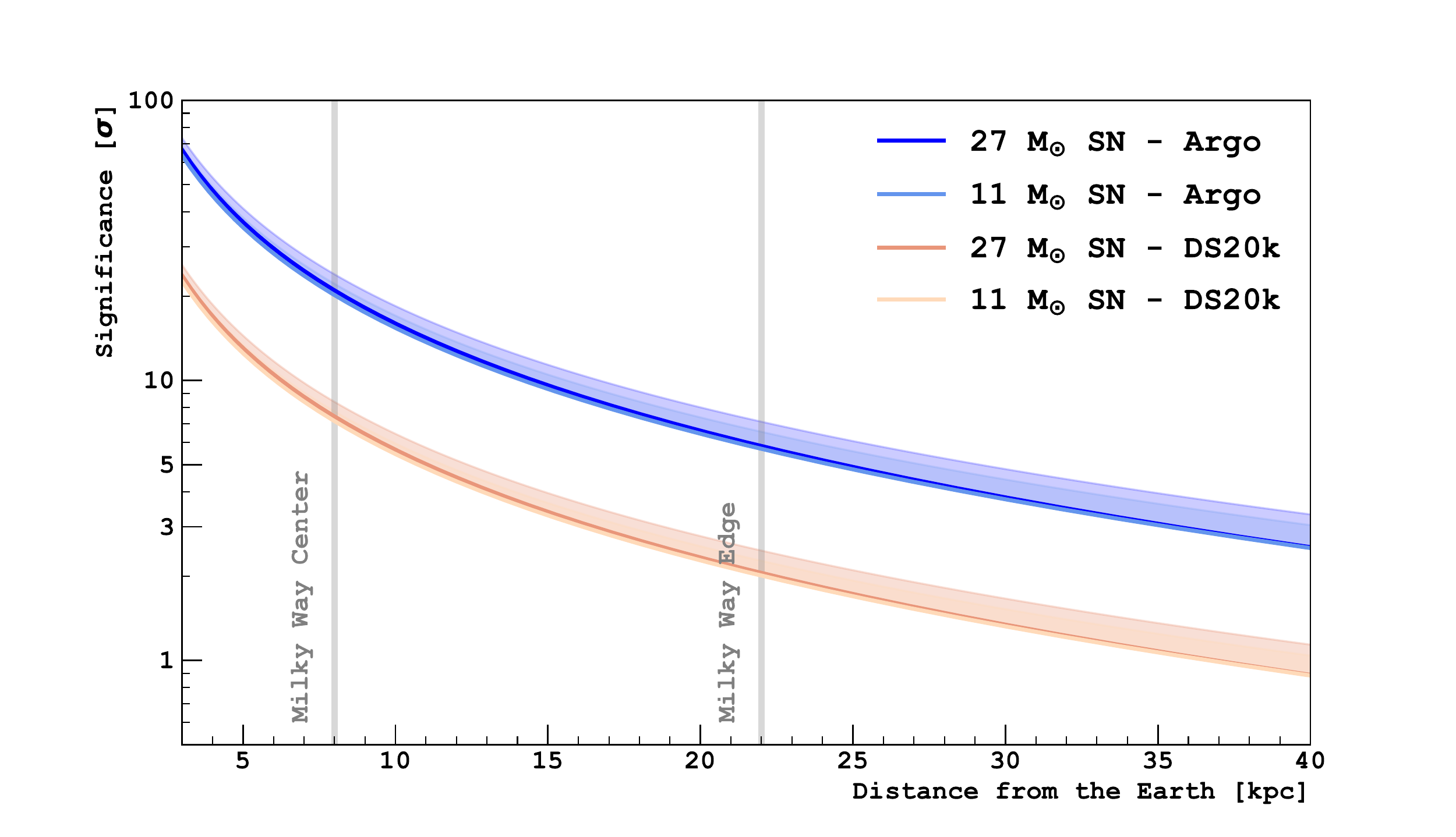}
   \caption{Top. \DSk\ and \Argo\ significance to 11-M$_{\odot}$ and 27-M$_{\odot}$ SNe (top) and to its neutronization burst only (bottom), as a function of the distance, assuming the standard background hypothesis (solid line) and (band)  lower contamination of $^{39}$Ar up to a factor of 10 less. Vertical lines represent the distance from the Earth of the Milky Way center and farthest edge, and of Large (LMC) and Small (SMC) Magellanic Clouds.  }
    \label{fig:significance}
\end{figure}

The  background expected  in  \DSk\ and \Argo\    can be assumed to be constant in time and known with negligible uncertainty, as it will be measured with very high statistics before and after the SN burst. This allows to  estimate the median significance  using the Asimov approximation  for likelihood-based tests  \cite{Cowan:2010js}.  The significance for both the TPCs and both the 11-M$_{\odot}$ and 27-M$_{\odot}$ SN models, assuming the background rate from table \ref{tab:eventcomparison}, is shown with solid lines in  \reffig{significance}, as a function of the SN distance from the Earth. The \DSk\ discovery potential entirely covers distances up to the edge of the Milky Way, and  \Argo\ extends it up almost to the Small Magellanic Cloud. As shown by the bands in \reffig{significance}, the potential increases significantly by assuming  lower contamination of $^{39}$Ar, as suggested in the previous section, up to a factor of 10 less.

The statistics of SN burst induced events in \Argo, and consequently its discovery sensitivity, is comparable to that of DUNE \cite{Abi:2020lpk}: the smaller active mass of  \Argo\, two orders of magnitude lower than DUNE, is compensated by the higher \CEnNS\ cross section, the lower energy threshold, and the possibility of observing all neutrino flavors. Compared with LXe target experiments, which also relay on the \CEnNS\  interaction channel, \DSk\ (\Argo) has slightly higher discovery sensitivity than XENONnT and LZ (DARWIN) \cite{PhysRevD.94.103009}.  An  extensive comparison with LXe, Cherenkov and liquid scintillator experiments can be found in  \cite{Kharusi:2020ovw}.

As for the neutronization burst only, \DSk \ can detect it as far as 10~kpc with a confidence level of 5~$\sigma$, and \Argo \ can extend it to $\sim$22~kpc, a distance equivalent to the farthest edge of the Milky Way from the Earth.  In this case, the significance, shown in \reffig{significance},  is similar for the two analyzed 11-M$_{\odot}$ and 27-M$_{\odot}$ SN models, as the number of events expected in the neutronization burst differs by only $\sim$10\%. 

 The detection sensitivity can be compared with the most recent determination of the expected SN core-collapse rate,  namely one event every 50 years within 30~kpc inside the Milky Way, and one event every 30 years within  3~Mpc, which includes  the Local Group \cite{Cappellaro:1999qy, Li:2000fp, Botticella:2016rul}.   
The SN  rate could be higher, with an upper limit of $\sim$20\%, because of  ``failed'' SNe, \textit{i.e.} core collapses of massive stars that form a black hole without  or with a little optical signature \cite{Gerke:2014ooa}. From a theoretical point of view,  the progenitor star may go through a neutronization stage with neutrino emission, during the collapse in the black hole.  Although GADMC TPCs have the potential to observe neutrinos emitted from failed SNe, their contribution was not included in this work and will be the subject of a future sensitivity study.

\DSk\ and \Argo, besides their use as counting experiments, can also provide information on the time and energy evolution of the neutrino flux. Simulations are performed, using a toy Monte Carlo approach, by applying on an event-by-event basis the detector response described in section \ref{sec:signal} to the interaction rate, obtained from the convolution of the neutrino flux from Garching simulations with the \CEnNS\ cross-section (eq. \ref{eq:cenns}). 

\begin{figure}
   \centering
    \includegraphics[width=1.0\linewidth]{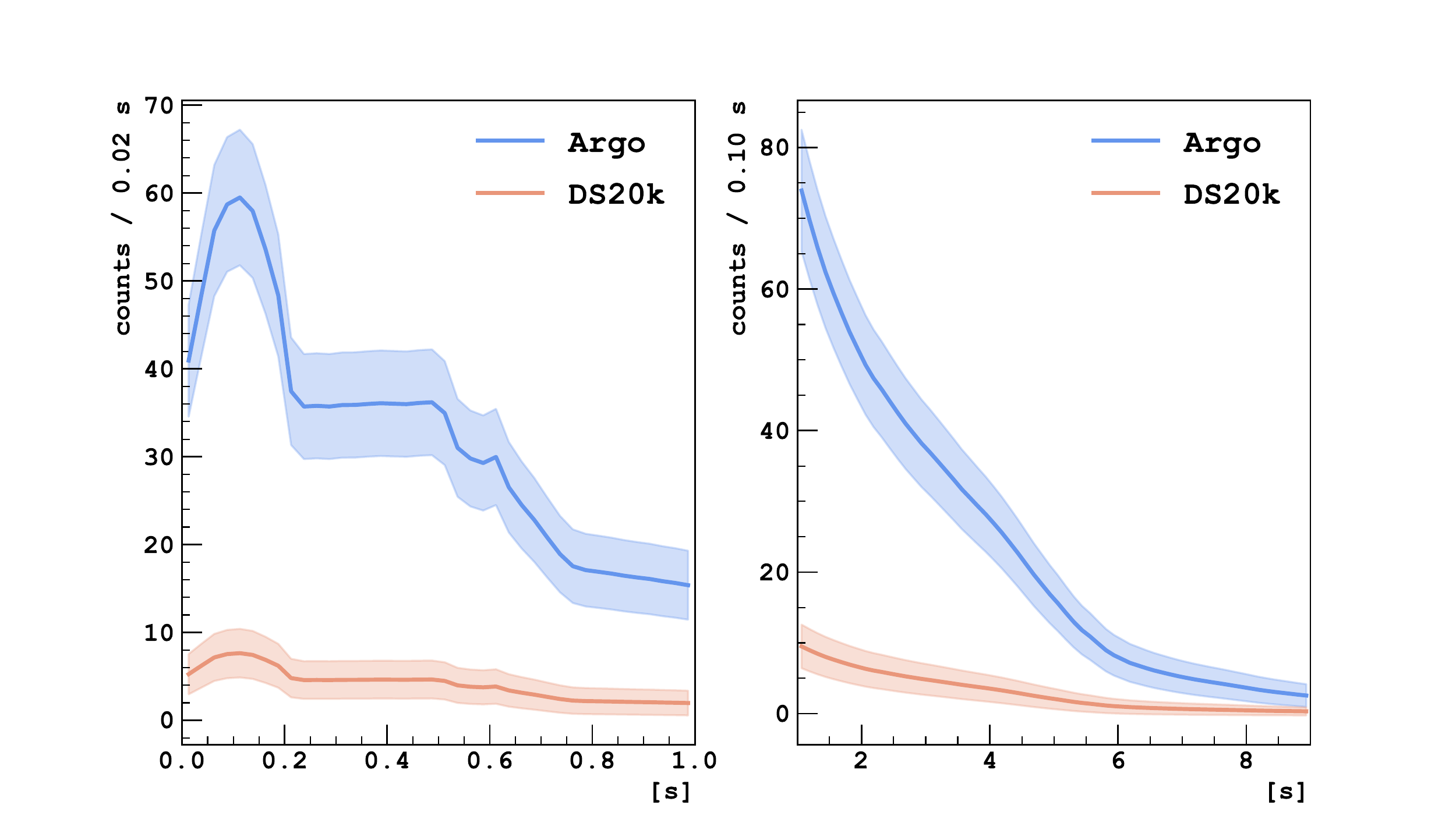}
   \caption{Time profile of neutrinos from the accretion (left) and cooling (right) phases of a 27-M$_{\odot}$ SN at 10 kpc distance,  as detected by \DSk\ and \Argo. The bands represent the statistical uncertainty.   }
    \label{fig:time_profile}
\end{figure}

The simulated time evolution of the accretion and cooling phases, as detected with  \DSk\ and \Argo, is shown in \reffig{time_profile} for a 27-M$_{\odot}$ SN at 10~kpc. The  energy window is limited to [3, 100]~\nel, where the background is almost entirely suppressed. The statistical error bands of the signal events are evaluated with respect to  the sampling  of 20 and 100 ms for the two phases, respectively.  The detector time responses of \DSk\ and \Argo, dominated by the associated electron drift times, are included in the simulations.  
It is worth highlighting that the statistics expected in \Argo, together with the time resolution, allows to distinguish the temporal structures that characterize the different SN  phases, and therefore to better constrain the models.

\begin{figure}
   \centering
\includegraphics[width=1.0\linewidth]{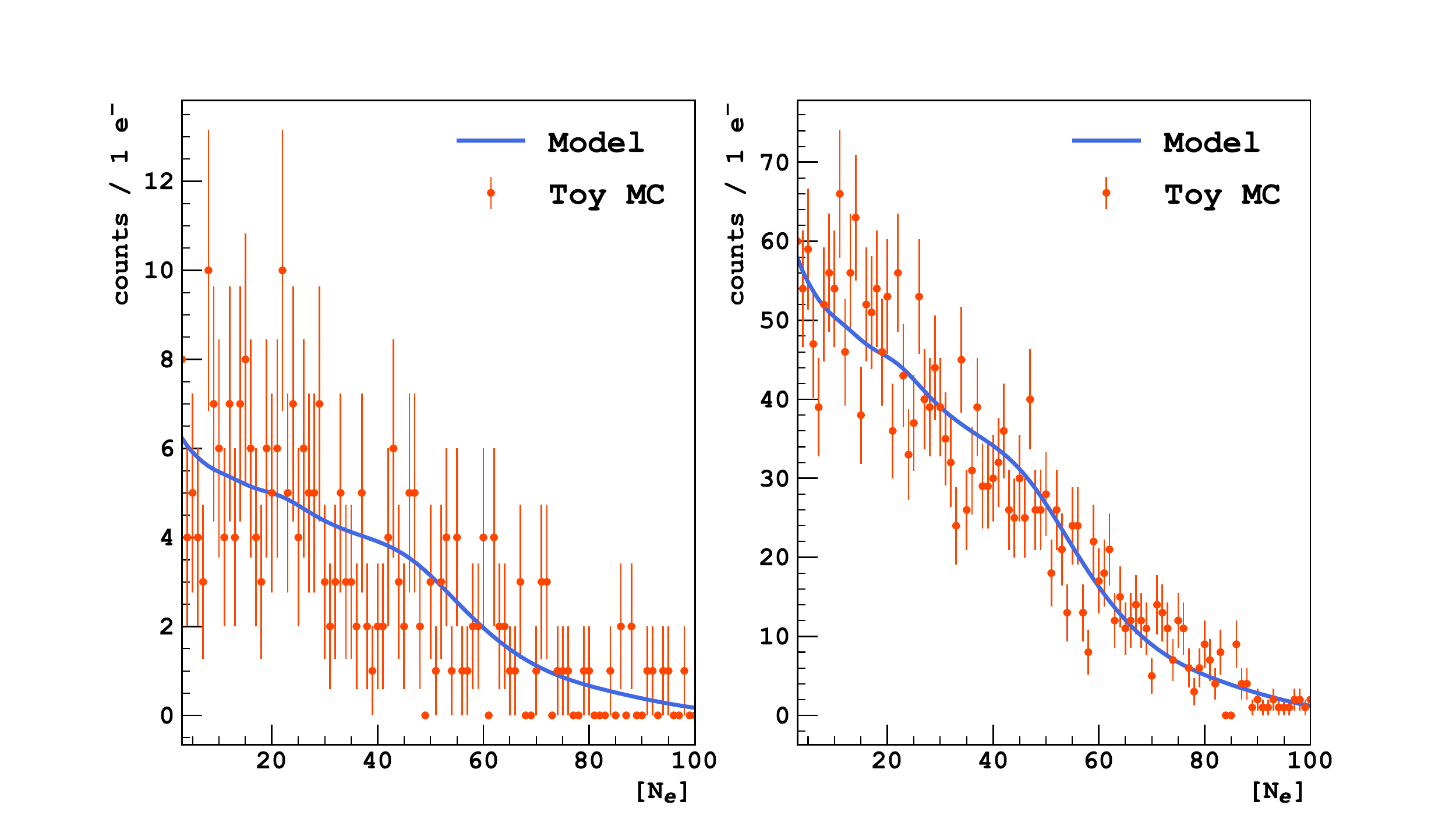}
      
   \caption{Examples of fit of two toy Monte Carlo neutrino interaction samples in \DSk\ (left) and \Argo\ (right), generated in the  [0.02, 8]~s  time range, corresponding to the   accretion and cooling phases from  a 27 M$_{\odot}$ SN burst at 10~kpc.   }
    \label{fig:toys}
\end{figure}

Examples of toy Monte Carlo samples in the \nel\ observable for the accretion phase only and  for all the SN phases but neutronization burst are shown in \reffig{toys}. These samples were produced for \Argo, assuming the neutrino flux from a 10~kpc distant 27-M$_{\odot}$ SN. From now on, we consider only this SN model for the following sensitivity study.  

The energy spectrum of the sum of all the SN emitted neutrino components can be parametrized  with  \cite{Keil_Raffelt_Janka_2003}

\begin{equation} \label{eq:nusp}
f(E_\nu) = \frac{\xi}{4{\pi}D^2} \frac{{(\alpha_T+1)^{\alpha_T+1}}{E_\nu}^{\alpha_T}  {e^{-\frac{E_\nu (\alpha_T+1)}{\langle E_\nu \rangle} } }}{{\langle E_\nu \rangle}^{{\alpha_T+1}}{\Gamma(\alpha_T+1)} }, 
\end{equation}

\noindent where $E_{\nu}$ is the neutrino energy, $\xi$ and $\langle E_\nu \rangle$ are the  total and mean SN neutrino energies emitted via neutrinos, respectively, $\alpha_T$ the so-called pinching parameter, $D$    the distance to the SN, and $\Gamma$ the Euler  gamma function. The spectrum in the neutronization burst can be approximated assuming $\alpha_T$=3.0, and with $\alpha_T$=2.3  in the accretion phase, where the neutrino emission starts to have a thermal spectrum.  In the cooling phase, the neutrino emission is close to having a Maxwell-Boltzmann distribution, where $\alpha_T$=2.0. 

The parametrized flux in eq. \ref{eq:nusp},  convoluted with the \CEnNS\ cross-section and the detector response, is used  to fit  toy Monte Carlo samples, in order  to assess the \DSk\ and \Argo\ sensitivities to  the  total and mean SN neutrino energies. Because of the non-normal fluctuations in the detector response, especially when \nel\  is close to the detector threshold (3~\nel), the convolution with the detector response is performed using a migration matrix, which transforms   nuclear  recoil energy into the \nel\ response function. This accounts also for the \nel\ fluctuations as discussed in section \ref{sec:signal}. Examples of fits of toy Monte Carlo samples are shown in \reffig{toys}. 

We have analyzed the two previously mentioned cases: the cooling phase only, and the full SN spectrum, excluding the neutronization burst. This choice is motivated by the good approximation of eq.~\ref{eq:nusp} with the accretion phase spectrum, assuming $\alpha_T$=2.3, and the similar $\alpha_T$ value between the accretion and the cooling phase. For the latter case, as the cooling phase provides a larger  statistics with respect to the accretion one, we assume $\alpha_T$ fixed to 2.0.  The statistics from the neutronization burst only is too low to allow for a spectral fit. In addition, as already discussed, the pinching parameter  is too different from the other phases to allow for an overall approximation with a unique $\alpha_T$ value.

\begin{figure}
   \centering
  \includegraphics[width=1.0\linewidth]{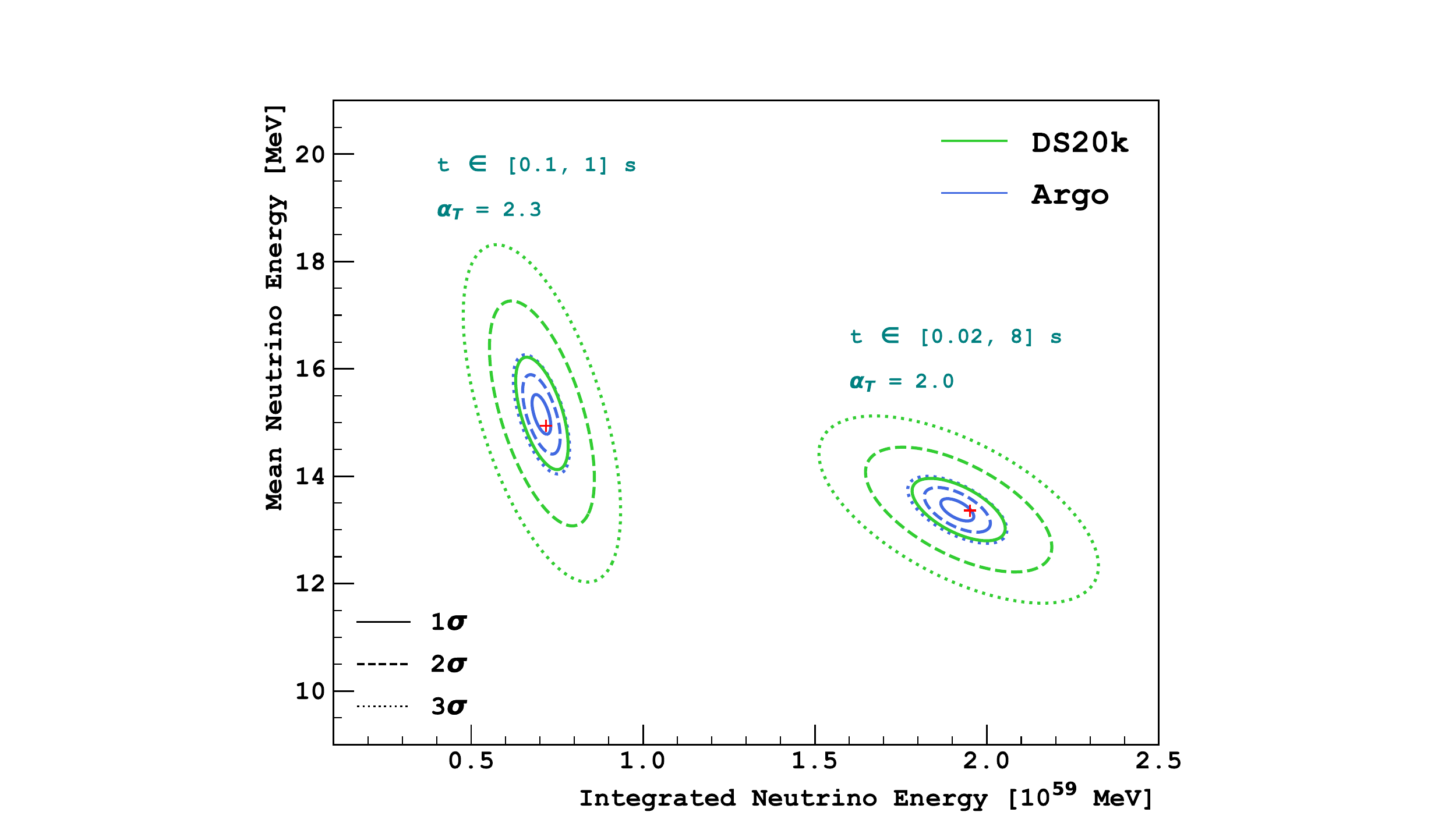}
   \caption{\DSk\ and \Argo\ sensitivities to mean and integrated neutrino energies of a 27 M$_{\odot}$ SN burst at 10~kpc in the [0.1, 1]~s  and [0.02, 8]~s. The two parameters are obtained by fitting 5$\times$10$^4$ toy MC samples with $\alpha_{T}$ equals to 2.3 and 2.0, with respect to each time range. Red crosses represent the true values from the Garching simulation. }
    \label{fig:reco}
\end{figure}

The sensitivities to $\langle E_\nu \rangle$ and $\xi$ in the accretion only and accretion+cooling phases are evaluated for both \DSk\ and \Argo. In each analyzed case, we have produced and fitted 5$\times$10$^4$ samples and derived the significance bands for 1, 2, and 3-$\sigma$ computed from the distribution of the best values from the fit. The results are shown in 
\reffig{reco}, together with the true values extracted from the original Garching simulations.

Both the experiments are able to reconstruct $\langle E_\nu \rangle$ and $\xi$ within 1-$\sigma$, even if a  systematic shift between true and reconstructed best values is present due to the parametrization approximation and the non-normal response of the detector.  The total neutrino energy is reconstructed at 3-$\sigma$ level  by \Argo\ (\DSk) with an accuracy of about 11\% (32\%) in the accretion-only and 7\% (21\%)  summing the contributions from both  accretion and cooling phases. For what concerns the mean energy,  \Argo\ has a 3-$\sigma$ level accuracy at 7\% in the accretion phase only, and at 5\% including also the cooling one. For the same parameter, \DSk\ can provide an accuracy of 21\% and 13\%, respectively. It is important to stress that the two parameters, as clearly visible in \reffig{reco},  are anti-correlated, with a  measured Pearson correlation coefficient of about -0.6 for all the analyzed cases.

% !TEX root = main.tex

\section{Conclusion and outlook}
\label{sec:sum}

\DSk\ and \Argo, with fiducial target masses of $\sim$50~t and $\sim$360~t, respectively, can detect neutrinos from SN burst via the flavor-insensitive \CEnNS\ channel, with an energy threshold of 0.46~keV$_{nr}$. Such a low analysis energy threshold can be achieved thanks to the $\sim$20\% accuracy in detecting single ionization electrons, as already demonstrated by \DSf. 
 
The low energy threshold, the resolution in the single-electron response, and the relatively low mass of the argon nucleus, compared to xenon, which kinematically extends the nuclear recoil spectrum at higher energies, allow the  LAr TPCs to achieve good accuracies in the reconstruction of average and integrated SN-emitted neutrino energies. Moreover, the time evolution of the SN burst can be investigated with 1.1~ms and 1.6~ms resolutions for \DSk\ and \Argo, respectively.
 
The discovery potential of a SN was also evaluated, demonstrating that \DSk\ can explore 11-M$_{\odot }$ and larger SNe up to the Milky Way edge, and  \Argo\ up to the Small Magellanic Cloud. Both \DSk\ and \Argo\ detectors are also sensitive to neutrinos from the 11-M$_{\odot }$ neutronization burst  up to the Milky Way center and edge, respectively.  These results take into account the most conservative predictions of $^{39}$Ar contamination. As already discussed, recent investigations from the DarkSide Collaboration suggest that the $^{39}$Ar contamination, intrinsic to underground argon, could be lower than the DarkSide-50 measured one, leading to a potential further improvement of the \DSk\ and  \Argo\ sensitivities.

The flavour-blind measurement from  \DSk\  and \Argo\ could be combined with the flavour-sensitive measurements of other neutrino detectors to provide another input into the triangulation of the positions of SNe, to be carried out by the SuperNova Early Warning System 2.0 (SNEWS2.0) \cite{Kharusi:2020ovw},  and to constrain the neutrino mass ordering by comparing electron flavor neutrino flux with the flavor-blind one from the neutronization stage. Furthermore, a measurement of the entire neutrino flux from the neutronization burst allows for the determination of  the SN distance within a precision of  $\sim$5\% \cite{Kachelriess:2004ds},  making detections via  \CEnNS\  channel   a potential  standard candle for distance measurements in the Milky Way Galaxy. Sensitivity studies for each of these physics measurements with the flavour-blind GADMC LAr TPC neutrino detection will be carried out in the future.

\acknowledgments
We are grateful to Prof. Alessandro Mirizzi, who provided the input fluxes for this study and invaluable sustain in our discussions on the supernova explosion mechanism. We also thank Mariangela Settimo for the useful comments. 
The DarkSide Collaboration would like to thank LNGS and its staff for invaluable technical and logistical support. This report is based upon work supported by the U. S. National Science Foundation (NSF) (Grants No. PHY-0919363, No. PHY-1004054, No. PHY-1004072, No. PHY-1242585, No. PHY-1314483, No. PHY- 1314507, associated collaborative grants, No. PHY-1211308, No. PHY-1314501, No. PHY-1455351 and No. PHY-1606912, as well as Major Research Instrumentation Grant No. MRI-1429544), the Italian Istituto Nazionale di Fisica Nucleare (Grants from Italian Ministero dell'Istruzione, Universit\`a, e Ricerca Progetto Premiale 2013 and Commissione Scientific Nazionale II), the Natural Sciences and Engineering Research Council of Canada, SNOLAB, and the Arthur B. McDonald Canadian Astroparticle Physics Research Institute. We acknowledge the financial support by LabEx UnivEarthS (ANR-10-LABX-0023 and ANR-18-IDEX-0001), the S\~ao Paulo Research Foundation (Grant FAPESP-2017/26238-4), and the Russian Science Foundation Grant No. 16-12-10369. The authors were also supported by the ``Unidad de Excelencia Mar\'{\i}a de Maeztu: CIEMAT - F\'{\i}sica de part\'{\i}culas'' (Grant MDM2015-0509), the Polish National Science Centre (Grant No. UMO-2019/33/B/ST2/02884), the Foundation for Polish Science (Grant No. TEAM/2016-2/17), the International Research Agenda Programme AstroCeNT (Grant No. MAB/2018/7) funded by the Foundation for Polish Science from the European Regional Development Fund, the Science and Technology Facilities Council, part of the United Kingdom Research and Innovation, and The Royal Society (United Kingdom). I.F.M.A is
supported in part by Conselho Nacional de Desenvolvimento Científico e Tecnol\'ogico (CNPq).  We also wish to acknowledge the support from Pacific Northwest National Laboratory, which is operated by Battelle for the U.S. Department of Energy under Contract No. DE-AC05-76RL01830.

\bibliographystyle{JHEP}
\bibliography{references}

%\begin{thebibliography}{99}
%\input{references.bib}

%\end{thebibliography}
\end{document}